\documentclass[pre,tighten]{revtex4}

\usepackage{amssymb,amsmath}

\usepackage{graphicx}

\newcommand\beq{\begin{equation}}
\newcommand\eeq{\end{equation}}
\newcommand\beqa{\begin{eqnarray}}
\newcommand\eeqa{\end{eqnarray}}

\newcommand{\al}{\alpha}

\begin{document}

\title{High-degree collisional moments of inelastic Maxwell mixtures. Application to the homogeneous cooling and uniform shear flow states}


\author{Constantino S\'anchez Romero\footnote[1]{Electronic address: constan@unex.es}}
\affiliation{Departamento de F\'{\i}sica, Universidad de Extremadura, Avda. de Elvas s/n, E-06006 Badajoz, Spain}
\author{Vicente Garz\'{o}\footnote[1]{Electronic address: vicenteg@unex.es;
URL: http://www.unex.es/eweb/fisteor/vicente/}}
\affiliation{Departamento de F\'{\i}sica and Instituto de Computaci\'on Ci\'ent\'ifica Avanzada Universidad (ICCAEx), Universidad de Extremadura, Avda. de Elvas s/n, E-06006 Badajoz, Spain}

\begin{abstract}
The Boltzmann equation for $d$-dimensional inelastic Maxwell models is considered to determine 
the collisional moments of second, third and fourth degree in a granular binary mixture. These collisional moments are exactly evaluated in terms of the velocity moments of the distribution function of each species when diffusion is absent (mass flux of each species vanishes). The corresponding associated eigenvalues as well as cross coefficients are obtained as functions of the coefficients of normal restitution and the parameters of the mixture (masses, diameters and composition). The results are applied to the analysis of the time evolution of the moments (scaled with a thermal speed) in two different nonequilibrium situations: the homogeneous cooling state (HCS) and the uniform (or simple) shear flow (USF) state. In the case of the HCS, in contrast to what happens for simple granular gases, it is shown that the third and fourth degree moments could diverge in time for given values of the parameters of the system. An exhaustive study on the influence of the parameter space of the mixture on the time behavior of these moments is carried out. Then, the time evolution of the second- and third-degree velocity moments in the USF is studied in the tracer limit (namely, when the concentration of one of the species is negligible). As expected, while the second-degree moments are always convergent, the third-degree moments of the tracer species can be also divergent in the long time limit.

\end{abstract}

\draft
\date{\today}
\maketitle

\section{Introduction}
\label{sec1}

It is well established that when granular matter is externally excited it can be modeled as a gas of \emph{inelastic} hard spheres (IHS). In the simplest version of the model, the spheres are assumed to be completely \emph{smooth} (i.e., with no rotational degrees of freedom) so that the inelasticity of collisions is characterized by a (positive) constant coefficient of normal restitution $\alpha\leq 1$. The case $\alpha=1$ corresponds to elastic collisions (molecular gases). In the low-density regime, the time evolution of the one-particle velocity distribution function is given by the Boltzmann kinetic equation properly adapted to account for the inelastic nature of collisions \cite{BP04}. Needless to say, the knowledge of the distribution function provides all the relevant information on the state of the gas at both microscopic and macroscopic levels.

However, the fact that the collision rate for hard spheres is proportional to the relative velocity of the two colliding particles hinders the search for solutions to the Boltzmann equation. In particular, this difficulty (which is also present for molecular gases) prevents the possibility of expressing the associated collisional moments of the Boltzmnn operator in terms of a finite number of velocity moments. This precludes for instance the derivation of \emph{exact} analytical results for the transport properties of the gas. For this reason, most of the analytical results derived for IHS are based on the truncation of a series expansion of the distribution function in powers of Laguerre (or Sonine) polynomials. In the case of elastic collisions, the above problem for the collisional kernel of hard spheres can be overcome by assuming that the particles interact via a repulsive potential inversely proportional to the fourth power of the distance (Maxwell molecules) \cite{TM80}.
For this interaction potential, the collision rate is independent of the relative velocity and so, any collisional moment of degree $k$ can be expressed in terms of velocity moments of a degree smaller than or equal to $k$ \cite{TM80}. Thanks to this property, nonlinear transport properties can be exactly obtained \cite{TM80,GS03} and, when properly reduced, they exhibit a good agreement with results derived for other interaction potentials.

In the case of granular gases (namely, when the collisions are inelastic), one can still introduce the so-called inelastic Maxwell models (IMM) (see for instance Refs.\ \cite{BCG00,CCG00,BK00} for some of the first papers where IMM were introduced). These models share with elastic Maxwell molecules the property that the collision rate is independent of the relative velocity but, on the other hand, their collision rules are the same as for IHS. Thus, although IMM cannot be represented by any interaction potential, its use allows one to get \emph{exact} analytical results of the \emph{inelastic} Boltzmann equation. In fact, IMM qualitatively keep the correct structure and properties of the nonlinear macroscopic equations and obey Haff's law \cite{Haff}. In any case, as Ref.\ \cite{E81} claims, one can introduce Maxwell models in the framework of the Boltzmann equation at the level of the cross section without any reference to a specific interaction potential. Recently \cite{KS22}, an inelastic \emph{rough} Maxwell model has been also introduced in the granular literature.

The simplifications introduced by IMM in the kernel of the Boltzmann collision operator has allowed, in some cases, the determination of the dynamic properties of granular gases without employing uncontrolled approximations. For this reason, the Boltzmann equation for IMM has received a great attention of physicists and mathematicians in the last few years, specially in the study of overpopulated high energy tails in homogeneous states \cite{EB02,BMP02,NK02a} and in the evaluation of the transport coefficients \cite{S02,GA05}. The existence of high energy tails of the Boltzmann equation is common for IHS and IMM; however the quantitative predictions of IMM differ from those obtained from IHS. A one-dimensional IMM has been also employed to study the two-particle velocity correlations \cite{CPM07}. It is important to remark that most of the problems analyzed in the context of IMM have been focused on simple (monocomponent) granular gases.
Much less is known in the case of inelastic Maxwell mixtures. For this sort of systems, Marconi and Puglisi have studied the high velocity moments in the free cooling \cite{MP02a} and driven \cite{MP02b} states for the one-dimensional case ($d=1$). For arbitrary dimensions and in the tracer limit, Ben-Naim and Krapivsky \cite{NK02} have analyzed the velocity statistics of an impurity in a uniform granular gas while the fourth cumulant of the velocity distribution in the homogenous cooling state (HCS) has been also obtained \cite{GA05}.


Beyond the second degree velocity moments (which are directly related with the transport properties), Garz\'o and Santos \cite{GS07} have computed all the third and fourth degree velocity moments of the Boltzmann collision operator for a monocomponent granular gas of IMM. In addition, the collisional rates associated with the isotropic velocity moments $\langle v^{2r} \rangle$ and the anisotropic moments $\langle v^{2r} v_i \rangle$ and $\langle v^{2r}\left( v_i v_j-d^{-1} v^2 \delta_{ij}\right) \rangle$ have been independently evaluated in Refs.\ \cite{EB02,BK03,SG12}. Here, $\langle h(\mathbf{v}) \rangle=\int d\mathbf{v} h(\mathbf{v}) f(\mathbf{v})$, where $h(\mathbf{v})$ is an arbitrary function of the velocity $\mathbf{v}$ and $f(\mathbf{v})$ is the one-particle velocity distribution function. All the above calculations have been performed for an arbitrary number of dimensions $d$. To the best of our knowledge, the above papers are the only works where the computation of high-degree collisional moments of IMM has been carried out.

On the other hand, as said before, the results for granular mixtures modeled as IMM are more scarce. In particular, given that most of the works have been focused on the computation of the transport coefficients, only the first- and second-degree collisional moments have been considered \cite{G03,GA05,GT10,GT11,GT12,GT15}. Thus, it would be convenient (specially for simulators) to extend the results displayed in Ref.\ \cite{GS07} for the third- and fourth-degree collisional moments to the realistic case of granular binary mixtures. This is the main objective of the present paper. However, due to the long and complex algebra involved in the general problem, here we will consider situations where the mean flow velocity $\mathbf{U}_r$ ($r=1,2$) of each species is equal to the mean flow velocity ${\bf U}$ of the mixture. This means that no diffusion processes are present in the mixture (i.e., $\mathbf{U}_1=\mathbf{U}_2=\mathbf{U}$). Although this limitation restricts the applicability of the present results to general nonequilibrium situations, they are still useful for contributing to the advancement in the knowledge of exact properties of IMM in some specific situations. Among the different problems, we can mention the relaxation of the third and fourth degree moments towards the HCS (starting from arbitrary initial conditions) and the study of the combined effect of shearing and inelasticity on the high-degree moments in a binary mixture under uniform shear flow (USF).

Some previous results derived in the HCS for IMM in the monodisperse case \cite{EB02, NK02} have shown that for $d\geq 2$ the (scaled) velocity distribution function $\phi(c)$ has a high-velocity tail of the form $\phi(c)\sim c^{-d-\beta(\alpha)}$ ($c$ being the (scaled) velocity of the particle). The exponent $\beta(\alpha)$ obeys a transcendental equation whose solution is always larger than four ($\beta(\alpha)>4$), except for the one-dimensional case ($d=1$) \cite{BMP02}. Consequently, for any value of $\alpha$ and $d \geq 2$, the corresponding (scaled) velocity moments of degree $k$ equal to or less than four tend towards well-defined values in the long-time limit (namely, they are always \emph{convergent}). An interesting issue is to explore whether or not the convergence of moments of degree $k\leq 4$ for the single gas case is also present for inelastic binary mixtures and, if so, to what extent. An indirect way of answering this question is through the knowledge of the high degree velocity moments (beyond the second ones) of the velocity distribution function of each species. These moments play a relevant role for instance in the high velocity region. Surprisingly, our results for binary mixtures show that the (anisotropic) third and fourth degree moments could diverge in time for given values of the parameters of the mixture. Therefore, in contrast to the findings for the monocomponent \emph{granular} gases for $d \geq 2$, only the (scaled) moments of degree equal to or smaller than 2 are always convergent in the HCS for arbitrary values of the parameters of the mixture. This is one of the main conclusions of the work.

Apart from the HCS, another interesting application of our results refers to the USF. For monocomponent granular gases, previous results \cite{SG07} have shown that, for a given value of the coefficient of restitution $\al$, the (scaled) symmetric fourth-degree moments diverge in time for shear rates larger than a certain critical value $a_c^*(\al)$. The value of $a_c^*(\al)$ decreases with decreasing $\al$ (increasing dissipation). Given that the analysis for general sheared binary mixtures is quite intricate, we consider here the limiting case where the concentration of one of the species is negligible (and so, it is present in tracer concentration). This limit allows one to express the moments of the tracer species in terms of the known moments of the excess gas. In particular, the knowledge of the second-degree moments provides the dependence of the temperature ratio on the parameters of the mixture. As occurs in the HCS, there is a breakdown of the energy equipartition; this behavior is produced here by the combined effect of both the shear rate and the inelasticity in collisions. In particular, in contrast to the HCS, we find a non-monotonic dependence of the temperature ratio on the (reduced) shear rate for given values of the coefficients of restitution. In addition, although the third-degree moments can be also divergent (as in the case of the HCS for mixtures and in contrast to the results reported for simple gases \cite{SG07}), surprisingly they become convergent for shear rates larger than a certain critical value.

The plan of the paper is as follows. In section \ref{sec2} the Boltzmann kinetic equation for inelastic Maxwell mixtures is presented. Next, the so-called Ikenberry polynomials \cite{TM80} $Y_{2p|i_1i_2\ldots i_q}(\mathbf{V})$ of degree $k=2p+q$ are introduced and their collisional moments $J_{2p|i_1i_2\ldots i_q}^{(rs)}$ with $k=2,3$ and 4 associated with the Boltzmann collision operators $J_{rs}[f_r,f_s]$ evaluated in section \ref{sec3}. Some technical details involved in the calculations are relegated to the Appendix \ref{appA}. The time relaxation problem of the (scaled) moments towards their asymptotic values in the HCS is studied in section \ref{sec4} while an study of the regions of the parameter space where the third- and fourth-degree moments can be divergent is presented in section \ref{sec5}. Section \ref{sec6} deals with the USF problem where we pay special attention to the second- and third-degree moments of the tracer species. Its time evolution is studied in section \ref{sec7}. We close the paper in section \ref{sec8} with a brief discussion of the results derived in this paper.

\section{Boltzmann kinetic equation for inelastic Maxwell mixtures}
\label{sec2}

We consider a granular binary mixture made of particles of diameters $\sigma_r$ and masses $m_r$ ($r=1,2$). In the absence of external forces and assuming molecular chaos, the one-particle velocity distribution function $f_r(\mathbf{r}, \mathbf{v};t)$ of species $r$ obeys the Boltzmann equation
\beq
\label{1.0}
\frac{\partial f_r}{\partial t}+\mathbf{v}\cdot \nabla f_r=\sum_{s=1}^2 J_{rs}[\mathbf{v}|f_r,f_s], \quad (r=1,2),
\eeq
where $J_{rs}[f_r,f_s]$ is the Boltzmann collision operator for collisions between particles of species $r$ and $s$. If the granular mixture is modeled as a gas of IHS then, to determine any collisional moment of $J_{rs}[f_r,f_s]$ one needs to know \emph{all} the degree moments of the distributions $f_r$ and $f_s$. This means that one has to resort to approximate forms of the distributions $f_r$ and $f_s$ to estimate the collisional moments of $J_{rs}$. Usually the lowest order in a Sonine polynomial expansion of these distributions is considered \cite{G19}. This problem is also present in the conventional case of molecular binary mixtures (elastic collisions). However, if one assumes that the collision rate of the two colliding spheres is constant (IMM), the collisional moments of the operator $J_{rs}[f_r,f_s]$ can be given in terms of velocity moments of the distributions $f_r$ and $f_s$ without knowing their explicit forms. This is the main advantage of using IMM instead of IHS.

The Boltzmann collision operator $J_{rs}[f_r,f_s]$ for IMM is \cite{G19}
\begin{equation}
J_{rs}\left[{\bf v}_{1}|f_r,f_s\right] =\frac{\omega_{rs}}{n_s\Omega_d}
\int d{\bf v}_{2}\int d\widehat{\boldsymbol{\sigma}}
\left[ \alpha_{rs}^{-1}f_r({\bf v}_{1}'')f_s({\bf v}_{2}'')-f_r({\bf v}_{1})f_s({\bf v}_{2})\right]
\;.
\label{1.1}
\end{equation}
Here,
\beq
\label{1.1.1}
n_r=\int d\mathbf{v} f_r(\mathbf{v})
\eeq
is the number density of species $r$, $\omega_{rs}$ is an effective collision
frequency (it can be seen as a free parameter of the model),  $\Omega_d=2\pi^{d/2}/\Gamma(d/2)$ is the total solid angle in $d$ dimensions, and $\alpha_{rs}\leq 1$ refers to the constant coefficient of restitution for $r$-$s$ collisions. In
addition, the double primes on the velocities denote the initial values $\{{\bf v}_{1}'',
{\bf v}_{2}''\}$ that lead to $\{{\bf v}_{1},{\bf v}_{2}\}$
following a binary collision:
\begin{equation}
\label{1.2}
{\bf v}_{1}''={\bf v}_{1}-\mu_{sr}\left( 1+\alpha_{rs}
^{-1}\right)(\widehat{\boldsymbol{\sigma}}\cdot {\bf g})\widehat{\boldsymbol
{\sigma}},
\quad {\bf v}_{2}''={\bf v}_{2}+\mu_{rs}\left(
1+\alpha_{rs}^{-1}\right) (\widehat{\boldsymbol{\sigma}}\cdot {\bf
g})\widehat{\boldsymbol{\sigma}}\;,
\end{equation}
where $\mu_{rs}=m_r/(m_r+m_s)$, ${\bf g}={\bf v}_1-{\bf v}_2$ is the relative velocity of the colliding pair and $\widehat{\boldsymbol{\sigma}}$ is a unit vector directed along the centers of the two colliding spheres.

Apart from the densities $n_r$, the granular temperature $T$ is defined as
\beq
\label{1.2.2}
T=\sum_{r=1}^2\; x_r T_r,
\eeq
where $x_r=n_r/n$ is the concentration or mole fraction of species $r$ ($n=n_1+n_2$ is the total number density) and
\beq
\label{1.2.3}
T_r=\frac{1}{d n_r} \int d\mathbf{v}\; m_r V^2 f_r(\mathbf{v})
\eeq
is the partial temperature of species $r$. In Eq.\ \eqref{1.2.3}, we have introduced the peculiar velocity $\mathbf{V}=\mathbf{v}-\mathbf{U}$, $\mathbf{U}$ being the mean
flow velocity defined as
\begin{equation}
\label{1.3}
\rho \mathbf{U}=\sum_{r=1}^2 \rho_r \mathbf{U}_r=\sum_{r=1}^2\int  d\mathbf{v} m_r\mathbf{v}f_r(\mathbf{v}).
\end{equation}
Here, $\rho_r=m_r n_r$ is the mass density of species $r$ and $\rho=\rho_1+\rho_2$ is the total mass density. The second identity in Eq.\ \eqref{1.3} defines the partial mean flow velocities $\mathbf{U}_r$. In addition, the mass flux of species $r$ is given by $\mathbf{j}_r=\rho_r \left(\mathbf{U}_r-\mathbf{U}\right)$. As said in section \ref{sec1}, for the sake of simplicity, we will assume in this paper that the mass fluxes vanish (i.e., $\mathbf{U}_r=\mathbf{U}$).

To evaluate the collisional moments of the Boltzmann operator $J_{rs}[f_r,f_s]$, a useful identity for an arbitrary function $h({\bf v})$ is
\begin{equation}
\label{1.3.1}
\int d\mathbf{v}_1 h(\mathbf {v}_1)J_{rs}[\mathbf{v}_1|f_r,f_s]=\frac{\omega_{rs}}{n_s\Omega_d} \int d\mathbf{v}_1\,\int d\mathbf{v}_2f_r(\mathbf{v}_1)f_s(\mathbf{v}_2) \int
d\widehat{\boldsymbol{\sigma}}\,\left[h(\mathbf{v}_1')-h(\mathbf{v}_1\right],
\end{equation}
where
\begin{equation}
\label{1.3.2}
\mathbf{v}_1'=\mathbf{v}_1-\mu_{rs}(1+\alpha_{rs})( \widehat{\boldsymbol{\sigma }}\cdot \mathbf{g})\widehat{\boldsymbol{\sigma}}
\end{equation}
denotes the post-collisional velocity.


\subsection{Ikenberry polynomials}

In the case of Maxwell models (both elastic and inelastic), it is convenient to introduce the Ikenberry
polynomials \cite{TM80} $Y_{2p|i_1i_2\ldots i_q}(\mathbf{V})$ of degree $k=2p+q$. The Ikenberry polynomials are defined as $Y_{2p|i_1i_2\ldots i_q}(\mathbf{V})=V^{2p}Y_{i_1i_2\ldots
i_q}(\mathbf{V})$. Here, as noted in Ref.\ \cite{GS07}, the polynomial $Y_{i_1i_2\ldots i_q}(\mathbf{V})$ is obtained by subtracting from
$V_{i_1}V_{i_2}\ldots V_{i_q}$ that homogeneous symmetric polynomial of degree $q$ in the components of
$\mathbf{V}$ such as to annul the result of contracting the components of $Y_{i_1i_2\ldots i_q}(\mathbf{V})$ on
any pair of indices. The polynomials functions $Y_{2p|i_1i_2\ldots i_q}(\mathbf{V})$ of degree smaller than or
equal to four are
\begin{equation}
\label{1.4}
 Y_{0|0}(\mathbf{V})=1,\quad Y_{0|i}(\mathbf{V})=V_i,
\end{equation}
\begin{equation}
\label{1.5}
Y_{2|0}(\mathbf{V})=V^2,\quad Y_{0|ij}(\mathbf{V})=V_i V_j-\frac{1}{d}V^2\delta_{ij},
\end{equation}
\begin{equation}
\label{1.6}
Y_{2|i}(\mathbf{V})=V^2 V_i,\quad
Y_{0|ijk}(\mathbf{V})=V_i V_j V_k-
\frac{1}{d+2}V^2\left(V_i\delta_{jk}+V_j\delta_{ik}+V_k
\delta_{ij}\right),
\end{equation}
\begin{equation}
\label{1.7}
Y_{4|0}(\mathbf{V})=V^4,\quad
Y_{2|ij}(\mathbf{V})=V^2\left(V_i V_j-\frac{1}{d}V^2\delta_{ij}\right),
\end{equation}
\begin{eqnarray}
\label{1.8}
Y_{0|ijk\ell}(\mathbf{V})&=&V_i V_j V_k V_\ell-\frac{1}{d+4}V^2
\left(V_iV_j\delta_{k\ell}+V_iV_k\delta_{j\ell}+V_iV_\ell\delta_{jk}
+V_jV_k\delta_{i\ell}\right.\nonumber\\
& & \left.+V_jV_\ell\delta_{ik}+V_kV_\ell\delta_{ij}\right)
+\frac{1}{(d+2)(d+4)}V^4\left(\delta_{ij}\delta_{k\ell}+\delta_{ik}\delta_{j\ell}+\delta_{i\ell}\delta_{jk}
\right)\nonumber\\
&=&V_iV_jV_kV_\ell-\frac{1}{d+4}\left[Y_{2|ij}(\mathbf{V})\delta_{k\ell}+Y_{2|ik}(\mathbf{V})\delta_{j\ell}
+Y_{2|i\ell}(\mathbf{V})\delta_{jk} \right.\nonumber\\
&&\left. +Y_{2|jk}(\mathbf{V})\delta_{i\ell}
+Y_{2|j\ell}(\mathbf{V})\delta_{ik}
+Y_{2|k\ell}(\mathbf{V})\delta_{ij}\right]\nonumber\\
& & -\frac{1}{d(d+2)}Y_{4|0}(\mathbf{V})\left(\delta_{ij}\delta_{k\ell}+
\delta_{ik}\delta_{j\ell}+\delta_{i\ell}\delta_{jk}\right).
\end{eqnarray}
Let us introduce here the notation
\begin{equation}
\label{1.9}
M_{2p|i_1i_2\ldots i_q}^{(r)}=\int d{\bf V} Y_{2p|i_1i_2\ldots i_q}({\bf V}) f_r({\bf V}),
\end{equation}
\begin{equation}
\label{1.10}
J_{2p|i_1i_2\ldots i_q}^{(rs)}=\int d{\bf V} Y_{2p|i_1i_2\ldots i_q}({\bf V}) J_{rs}[f_r,f_s].
\end{equation}
Equation \eqref{1.9} gives the definition of the velocity moments of the distribution $f_r$ while Eq.\ \eqref{1.10} provides the definition of the collisional moments of the Boltzmann operator $J_{rs}$.

Note that $M_{0|0}^{(r)}=n_r$, $J_{0|0}^{(rs)}=0$ (conservation of mass), $M_{0|i}^{(r)}=0$ (since we have assumed
that $\mathbf{U}_r=\mathbf{U}$) and
\beq
\label{1.10.0}
M_{2|0}^{(r)}=d \frac{p_r}{m_r}=d \frac{n_r T_r}{m_r},
\eeq
where $p_r=n_r T_r$ is the partial pressure of species $r$.  Moreover,
\beq
\label{1.10.3}
M_{0|ij}^{(r)}=\frac{P_{r,ij}-p_r\delta_{ij}}{m_r},
\eeq
where
\begin{equation}
\label{1.10.1}
P_{r,ij}=\int d{\bf V}m_r V_i V_j f_r({\bf v})
\end{equation}
is the partial pressure tensor of species $r$ and
\beq
\label{1.10.4}
M_{2|i}^{(r)}=2\frac{q_{r,i}}{m_r},
\eeq
where
\begin{equation}
\label{1.10.2}
\mathbf{q}_r=\int d{\bf V}\frac{m_r}{2} V^2 {\bf V} f_r({\bf V})
\end{equation}
is the partial contribution to the total heat flux due to species $r$.

The remaining third degree moments $M_{0|ijk}^{(r)}$ and the moments of degree $k\geq 4$ are not directly related to hydrodynamic quantities. However, they provide indirect information on the velocity distribution function $f_r$.

\section{Collisional moments for inelastic Maxwell mixtures}
\label{sec3}

As mentioned in section \ref{sec1}, the main advantage of using IMM instead of IHS is that a collisional moment of degree $k$ of the Maxwell collision operator $J_{rs}[f_r,f_s]$ can be written as a bilinear combination of velocity moments of $f_r$ and $f_s$ of degree less than or equal to $k$. This result holds for elastic \cite{TM80,GS03} and inelastic \cite{G19} gases. Let us now display the explicit expressions for the collisional moments $J_{2p|i_1i_2\ldots i_q}^{(rs)}$ for $k=2p+q\leq 4$. Some technical details to obtain those collisional moments are provided in the Appendix \ref{appA}.

\subsection{Second degree collisional moments}

The second degree collisional moments were already evaluated in Refs.\ \cite{G03,GT10}. They are given by
\begin{equation}
\label{1.11}
{J}_{2|0}^{(rs)}= -\frac{\omega_{rs}}{4d n_s}(1+\beta_{rs})\left[(3-\beta_{rs})n_sM_{2|0}^{(r)}-
(1+\beta_{rs})n_rM_{2|0}^{(s)}\right],
\end{equation}
\begin{equation}
\label{1.12}
{J}_{0|ij}^{(rs)}=-\frac{\omega_{rs}}{2d(d+2)n_s}(1+\beta_{rs})\left[(2d+3-\beta_{rs})
n_sM_{0|ij}^{(r)}-(1+\beta_{rs})n_rM_{0|ij}^{(s)}\right],
\end{equation}
where we have introduced the auxiliary quantity
\begin{equation}
\label{1.12.0}
\beta_{rs}=2\mu_{sr}(1+\alpha_{rs})-1.
\end{equation}
For mechanically equivalent particles ($m_1=m_2$, $\sigma_1=\sigma_2$, $\al_{11}=\al_{22}=\al_{12}$), $\beta_{rs}=\al$.

The quantity $\zeta_{rs}$ measures the rate of change of the partial temperature $T_r$ due to collisions with particles of species $s$. It is defined as
\beq
\label{1.12.2}
\zeta_{rs}=-\frac{m_r}{d n_r T_r}{J}_{2|0}^{(rs)}.
\eeq
The total cooling rate $\zeta$ is
\begin{equation}
\label{1.12.1}
\zeta=T^{-1}\sum_{r=1}^2\; x_rT_r \zeta_r, \quad \zeta_r=\sum_{s=1}^2\; \zeta_{rs}.
\end{equation}
According to Eqs.\ \eqref{1.11} and \eqref{1.12.2}, the parameters $\zeta_{rs}$ can be written as
\begin{equation}
\label{2.3.1}
\zeta_{rs}=\frac{\omega_{rs}}{4d}(1+\beta_{rs})\left[3-\beta_{rs}-(1+\beta_{rs})\frac{m_r T_s}{m_s T_r}\right].
\end{equation}
Note that Eq.\ \eqref{1.11} yields the result
\beqa
\label{13.0}
\sum_{r=1}^2 \sum_{s=1}^2 m_r J_{2|0}^{(rs)}&=&-d n T \zeta= -\frac{\omega_{11}}{2}n_1 T_1 (1-\al_{11}^2)-\frac{\omega_{22}}{2}n_2 T_2 (1-\al_{22}^2)\nonumber\\
& &-\omega_{12}n_1
\mu_{21}\left(\mu_{21}T_1+\mu_{12} T_2\right)(1-\al_{12}^2).
\eeqa
For elastic collisions ($\al_{11}=\al_{22}=\al_{12}=1$), Eq.\ \eqref{13.0} shows that the total kinetic energy is conserved by collisions regardless of the values of the masses and diameters of the mixture. This is the expected result.

\subsection{Third degree collisional moments}

The evaluation of the third degree collisional moments is performed in the Appendix \ref{appA}.
The results are
\beqa
\label{1.13}
 {J}_{2|i}^{(rs)}&=&-\frac{1}{8d(d+2)}\frac{\omega_{rs}}{n_s}
 (1+\beta_{rs})\left\{
 \left[3\beta_{rs}^2-2(d+5)\beta_{rs}+10d+11\right]n_s M_{2|i}^{(r)}\right.\nonumber\\
 & &\left.-3(1+\beta_{rs})^2
 n_r M_{2|i}^{(s)}\right\},
\eeqa
\beqa
\label{1.14}
{J}_{0|ijk}^{(rs)}&=&-\frac{3}{4d(d+2)(d+4)}\frac{\omega_{rs}}{n_s}
(1+\beta_{rs})
\Big\{ \left[\beta_{rs}^2-2(d+3)\beta_{rs}+2d^2+10d+9)\right]
\nonumber\\
& & \times n_sM_{0|ijk}^{(r)}
-(1+\beta_{rs})^2n_rM_{0|ijk}^{(s)}\Big\}.
\eeqa
Equation (\ref{1.13}) is consistent with the expression derived in Ref.\ \cite{G03} when the mass flux $\mathbf{j}_r=\mathbf{0}$.

\subsection{Fourth degree collisional moments}

The calculation of the fourth degree collisional moments is more involved. The collisional moments
${J}_{4|0}^{(rs)}$ and ${J}_{2|ij}^{(rs)}$ can be written as
\beqa
\label{1.15}
{J}_{4|0}^{(rs)}&=&-\frac{1}{d(d+2)}\frac{\omega_{rs}}{n_s}(1+\beta_{rs})
\Bigg[\frac{(3-\beta_{rs})(3\beta_{rs}^2-6\beta_{rs}+8d+7)}{16}n_sM_{4|0}^{(r)}\nonumber\\
& &
-\frac{3}{16}(1+\beta_{rs})^3n_rM_{4|0}^{(s)}
-\frac{(1+\beta_{rs})(3\beta_{rs}^2-6\beta_{rs}-1)}{4}\frac{P_{r,ij}P_{s,ij}}{m_rm_s}\nonumber\\
& &
-\frac{(1+\beta_{rs})(3\beta_{rs}^2-6\beta_{rs}+4d+7)}{8}\frac{d^2p_rp_s}{m_rm_s}\Bigg],
\eeqa
\begin{eqnarray}
\label{1.16}
{J}_{2|ij}^{(rs)}
&=&\frac{1}{d(d+2)(d+4)}\frac{\omega_{rs}}{n_s}
(1+\beta_{rs})\left\{\frac{3}{4}(1+\beta_{rs})^3n_rM_{2|ij}^{(s)}\right.\nonumber\\
& & +
\frac{3\beta_{rs}^3-3(d+5)\beta_{rs}^2+\beta_{rs}(d^2+14d+25)-7d^2-31d-21}{4}n_s
M_{2|ij}^{(r)}\nonumber\\
& & +\frac{6\beta_{rs}^3-3(d+2)\beta_{rs}^2-2(d+7)\beta_{rs}+d-2}{4}
\frac{\left(P_{r,ik}P_{s,kj}+P_{r,jk}P_{s,ki}\right)}
{m_rm_s}\nonumber\\
& & +\frac{(1+\beta_{rs})(3\beta_{rs}^2+2d+5)}{4}\frac{dp_r}{m_rm_s}P_{s,ij}
\nonumber\\
& &
+\frac{(1+\beta_{rs})\left[3\beta_{rs}^2-3(d+4)\beta_{rs}+d^2+7d+9\right]}{4}
\frac{dp_s}{m_rm_s}P_{r,ij}\nonumber\\
& & -\frac{(1+\beta_{rs})[6\beta_{rs}^2-3(d+4)\beta_{rs}+d^2+9d+14]}{4}\frac{dp_rp_s}{m_rm_s}\delta_{ij}
\nonumber\\
& & \left.
-\frac{(1+\beta_{rs})[6\beta_{rs}^2-3(d+4)\beta_{rs}+d-2]}{2d}
\delta_{ij}\frac{P_{r,ij}P_{s,ij}}{m_rm_s}\right\},
\end{eqnarray}
The expression of ${J}_{0|ijk\ell}^{(rs)}$ is rather large. For the sake of concreteness, we will display here two representative collisional moments of this class: ${J}_{0|xxxx}^{(rs)}$ and ${J}_{0|xxxy}^{(rs)}$. They are given by
\begin{eqnarray}
\label{1.17} {J}_{0|xxxx}^{(rs)}
&=&\frac{1}{2d(d+2)(d+4)(d+6)}\frac{\omega_{rs}}{n_s} (1+\beta_{rs})\nonumber\\
&  & \times \Big\{\left[3\beta_{rs}^3
-(6d+27)\beta_{rs}^2+3\beta_{rs}(2d^2+16d+27)-4d^3-42d^2\right.\nonumber\\
& & \left.
-122d-81\right]
n_s{M}_{0|xxxx}^{(r)} +3(1+\beta_{rs})^3n_r{M}_{0|xxxx}^{(s)}\nonumber\\
& & +6\frac{3\beta_{rs}^2-3\beta_{rs}(d+4)+d^2+7d+9}{(d+4)}\frac{(1+\beta_{rs})}{m_rm_s}\nonumber\\
& & \times \Big[(d+2)^{-1}(2P_{r,ij}P_{s,ij}+d^2p_rp_s) -d(p_sP_{r,xx}+
p_rP_{s,xx})\nonumber\\
& &
-4P_{r,xk}P_{s,kx}+(d+4)
P_{r,xx}P_{s,xx}\Big]\Big\},
\end{eqnarray}
\begin{eqnarray}
\label{1.18} {J}_{0|xxxy}^{(rs)}
&=&\frac{1}{2d(d+2)(d+4)(d+6)}\frac{\omega_{rs}}{n_s} (1+\beta_{rs})\nonumber\\
&  & \times \Big\{\left[3\beta_{rs}^3
-(6d+27)\beta_{rs}^2+3\beta_{rs}(2d^2+16d+27)-4d^3-42d^2\right.\nonumber\\
& & \left.
-122d-81\right]
n_s{M}_{0|xxxy}^{(r)} +3(1+\beta_{rs})^3n_r{M}_{0|xxxy}^{(s)}\nonumber\\
& & +3\frac{3\beta_{rs}^2-3\beta_{rs}(d+4)+d^2+7d+9}{(d+4)}\frac{(1+\beta_{rs})}
{m_rm_s}
\nonumber\\
& & \times \Big[(d+4)\left(P_{r,xx}P_{s,xy}+P_{r,xy}P_{s,xx}\right)-
d(p_sP_{r,xy} +p_rP_{s,xy})\nonumber\\
& &
-2\left(P_{r,xk}P_{s,ky}+P_{r,yk}P_{s,kx}\right)\Big]
\Big\}.
\end{eqnarray}

In the case of mechanically equivalent particles, all the expressions reduce to known results for molecular ($\al=1$) three-dimensional ($d=3$) Maxwell gases \cite{TM80,GS03}. Regarding inelastic collisions ($\alpha <1$) and mechanically equivalent particles, the expressions for the collisional moments are consistent with the results reported in Ref.\ \cite{GS07} for monocomponent granular gases. Moreover, in the one-dimensional case ($d=1$, $\alpha_{rs} <1$), our results for granular binary mixtures of the isotropic collisional moments ${J}_{2|0}^{(rs)}$ and ${J}_{4|0}^{(rs)}$ agree with those previously reported by Marconi and Puglisi \cite{MP02a}. This shows the consistency of our expressions with known results published in the granular literature.


As said before, since the collision frequencies $\omega_{rs}$ can be seen as free parameters in the model, the expressions obtained in this section for the collisional moments apply regardless the specific choice of the above frequencies. Although $\omega_{rs}$ is independent of velocity, it can  depend on space an time through its dependence on density and temperature. On physical grounds, $\omega_{rs}\propto n_s$ since $n_r \omega_{rs}=n_s \omega_{sr}$. As in
previous works of granular mixtures \cite{GT10,GT15}, we will assume that
$\omega_{rs}\propto n_s T^\beta$, with $\beta\geq 0$. The case $\beta=0$ (a collision frequency independent of temperature) will be referred to as Model A while the case $\beta\neq 0$ (collision frequency monotonically increasing with temperature) will be called Model B. Model A is closer to the original model of Maxwell molecules for elastic gases while Model B, with $\beta=\frac{1}{2}$, is closer to hard spheres.

\section{Relaxation to the HCS}
\label{sec4}

The results derived in the preceding section can be applied to several interesting situations. In this paper, we will consider first the most basic problem in a granular mixture: the  time evolution of the moments of degree less than or equal to four (both isotropic and anisotropic) in the HCS. The HCS is a \emph{homogeneous} state where the granular temperature $T(t)$ monotonically decays in
time. In this case, the set of Boltzmann kinetic equations \eqref{1.0} for $f_1$ and $f_2$ becomes
\begin{equation}
\label{2.1}
\partial_t f_1(\mathbf{v},t)=J_{11}[\mathbf{v}|f_1,f_1]+J_{12}[\mathbf{v}|f_1,f_2],
\end{equation}
\beq
\label{2.1.0}
\partial_t f_2(\mathbf{v},t)=J_{22}[\mathbf{v}|f_2,f_2]+J_{21}[\mathbf{v}|f_2,f_1].
\eeq
In the HCS, the granular temperature $T(t)$ decreases in time due to collisional dissipation. A steady state can be achieved if some sort of thermostat (which injects energy into the system) is introduced in the system to compensate for the energy dissipated by collisions. Here, we will assume that the granular mixture is \emph{undriven} and hence, $T$ depends on time.

In the context of IMM, it has been proven for single granular gases \cite{BC03,BCT03} that, provided that
$f_r(\mathbf{v},0)$ $(r=1,2)$ has a finite moment of some degree higher than two, $f_r(\mathbf{v},t)$ asymptotically tends
toward a self-similar solution of the form
\begin{equation}
\label{2.2}
f_r(\mathbf{v},t)\to n_r [v_0(t)]^{-d} \phi_r(c), \quad c=v/v_0(t),
\end{equation}
where $\phi_r(c)$ is an  isotropic distribution of the scaled velocity $c$. However, the exact form of the distribution $\phi_r(c)$ is not known to date.

\begin{figure}[h]
\includegraphics[width=0.4\textwidth]{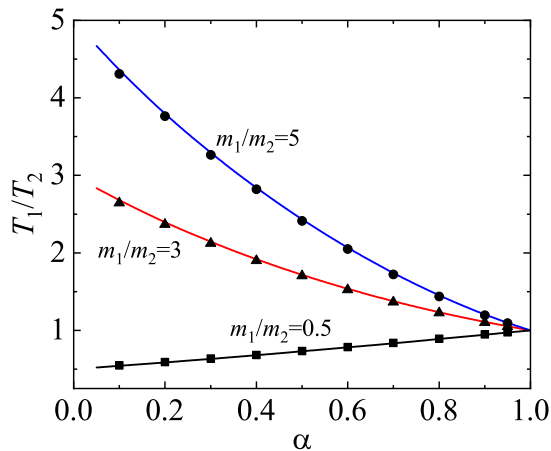}
\caption{Temperature ratio $T_1/T_2$ versus the (common) coefficient of restitution $\al$ for a three-dimensional binary mixture ($d=3$) with $\sigma_1/\sigma_2=1$, $x_1=\frac{1}{2}$, and three different values of the mass ratio: $m_1/m_2=5$ (circles), $m_1/m_2=5$ (triangles), and $m_1/m_2=0.5$ (squares). The lines correspond to the theoretical results obtained here for IMM while the symbols refer to the results obtained from the DSMC method for IHS \cite{ChGG22}. The theoretical lines have been obtained from the condition $\zeta_1=\zeta_2$.}
\label{fig1}
\end{figure}

At a hydrodynamic level, the only relevant balance equation is that of the temperature $T(t)$. Its time evolution equation can be easily obtained from the moments $M_{2|0}^{(1)}$ and $M_{2|0}^{(2)}$ and it is given by
\begin{equation}
\label{2.5.1}
\partial_t T=-\zeta T,
\end{equation}
where we have taken into account that the time evolution of the partial temperatures $T_r$ can be derived from the velocity moments $M_{2|0}^{(r)}$ as
\beq
\label{2.2.0}
\partial_t T_r=-\zeta_r T_r.
\eeq
Here, we recall that $\zeta_r=\sum_s \zeta_{rs}$ and $\zeta_{rs}$ is given by Eq.\ \eqref{2.3.1}. The time evolution of the temperature ratio $\gamma\equiv T_1(t)/T_2(t)$ follows from Eq.\ \eqref{2.2.0} as
\beq
\label{2.2.1}
\partial_t \ln \gamma=\zeta_2-\zeta_1.
\eeq
After a transient period, it is expected that the mixture achieves a hydrodynamic regime where all the time dependence of $f_r$ only occurs through the granular temperature $T(t)$. This necessarily implies that the three temperatures $T_1(t)$, $T_2(t)$ and $T(t)$ are proportional to each other and their ratios are all constant. This does not necessarily means that all three temperatures are equal (as in the case of elastic collisions) since the value of $\gamma$ must be obtained from Eq.\ \eqref{2.2.1}. In fact, in the hydrodynamic regime, $\gamma\equiv \text{const.}$ and so the condition of equal partial cooling rates [$\zeta_1(t)=\zeta_2(t)$] provides the dependence of the temperature ratio on the parameters of the mixture \cite{GD99}.

Figure \ref{fig1} shows the dependence of the temperature ratio $T_1/T_2$ on the (common) coefficient of restitution $\al_{rs}\equiv \al$ for a three-dimensional binary mixture ($d=3$) with $\sigma_1/\sigma_2=1$, $x_1=\frac{1}{2}$, and three different values of the mass ratio. To compare with the results obtained from IHS, we chose $\omega_{rs}$ to get the same $\zeta_{rs}$ [Eq.\ \eqref{2.3.1} for IMM] as that of IHS. In the case of IHS, the quantities $\zeta_{rs}$ are evaluated by approximating the distributions $f_r$ and $f_s$ by Maxwellian distributions defined at temperatures $T_r$ and $T_s$, respectively \cite{G03}. With this choice, in the case $\beta=\frac{1}{2}$, one achieves the expression
\beq
\label{w.1}
w_{rs}=\frac{2\pi^{(d-1)/2}}{\Gamma\left(\frac{d}{2}\right)}n_s \sigma_{rs}^{d-1}\left(\frac{2T_r}{m_r}+\frac{2T_s}{m_s}\right)^{1/2},
\eeq
where $\sigma_{rs}=(\sigma_r+\sigma_s)/2$. Theoretical results for IMM [with the choice \eqref{w.1}] are compared against Monte Carlo simulations carried out in Ref.\ \cite{ChGG22} for IHS. Figure \ref{fig1} highlights one of the most characteristic features of granular mixtures (as compared with molecular mixtures): the partial temperatures are different even in homogenous states. We observe that the departure of $\gamma$ from 1 (breakdown of energy equipartition) increases with increasing the disparity in the mass ratio. In general, the temperature of the lighter species is smaller than that of the heavier species. It is also important to remark the excellent agreement found between theory (developed for IMM) and computer simulations (performed for IHS), even for quite strong inelasticity.

\subsection{Eigenvalues for inelastic Maxwell mixtures}

\begin{figure}[ht]
\includegraphics[width=0.35\textwidth]{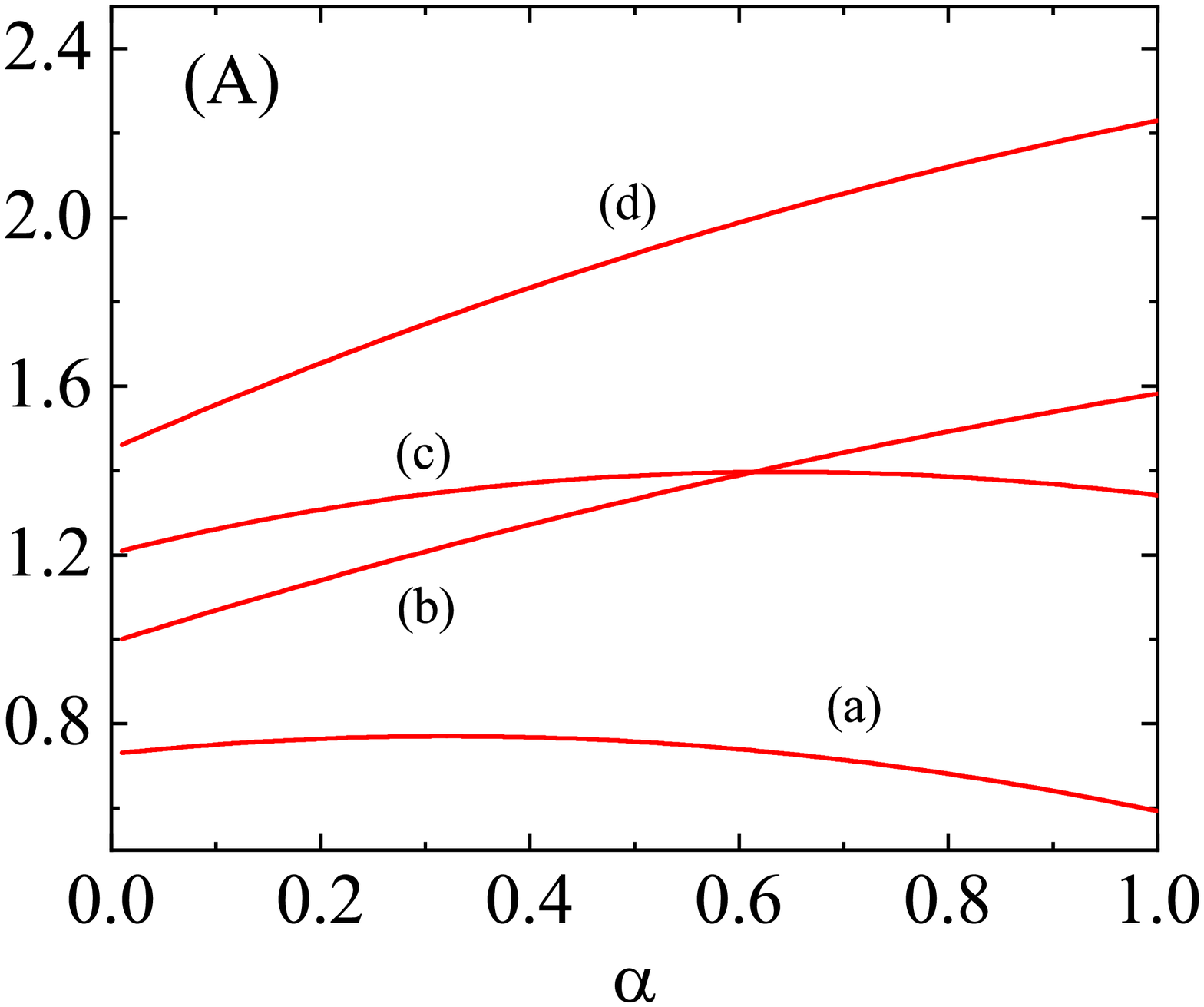}\hspace{5mm}
\includegraphics[width=0.35\textwidth]{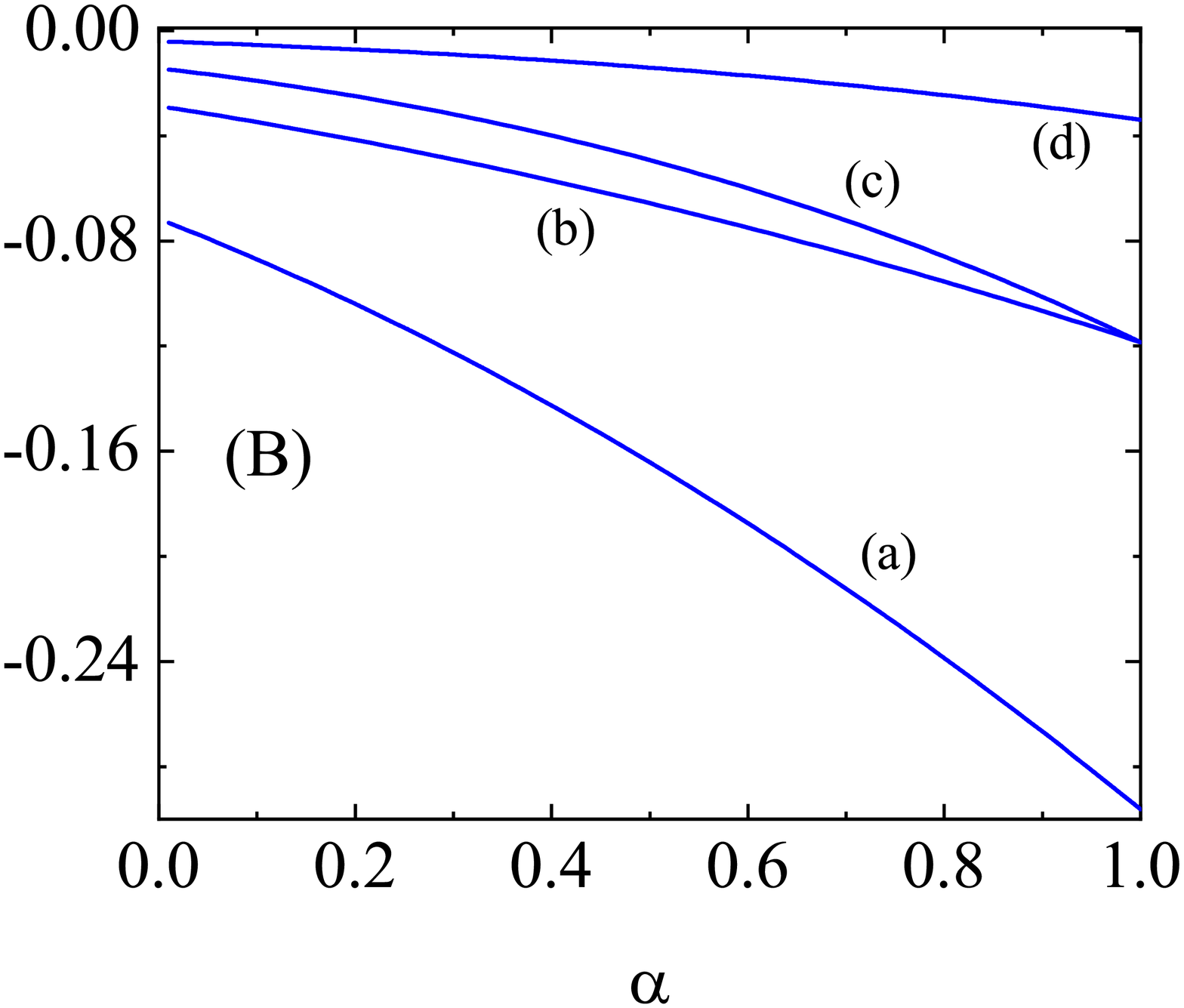}
\caption{Panel A: Dependence of the eigenvalues $\nu_{2|0}^{(11)}$ [defined by Eq.\ \eqref{b1}] (a), $\nu_{0|ij}^{(11)}$ [defined by Eq.\ \eqref{b2}] (b), $\nu_{2|i}^{(11)}$ [defined by Eq.\ \eqref{b3}] (c), and $\nu_{0|ijk}^{(11)}$ [defined by Eq.\ \eqref{b4.0}] (d) on the (common) coefficient of restitution $\al$ for a three-dimensional binary mixture constituted by particles of the same mass density [$m_1/m_2=(\sigma_1/\sigma_2)^3$] with $x_1=\frac{1}{2}$ and $m_1/m_2=2$. Panel B: The same as panel A for the eigenvalues $\nu_{2|0}^{(12)}$ [defined by Eq.\ \eqref{b1}] (a),
$\nu_{0|ij}^{(12)}$ [defined by Eq.\ \eqref{b2.0}] (b),
$\nu_{2|i}^{(12)}$ [defined by Eq.\ \eqref{b4}] (c), and $\nu_{0|ijk}^{(12)}$ [defined by Eq.\ \eqref{b4.1}] (d).
\label{fig2}}
\end{figure}

\begin{figure}[ht]
\includegraphics[width=0.35\textwidth]{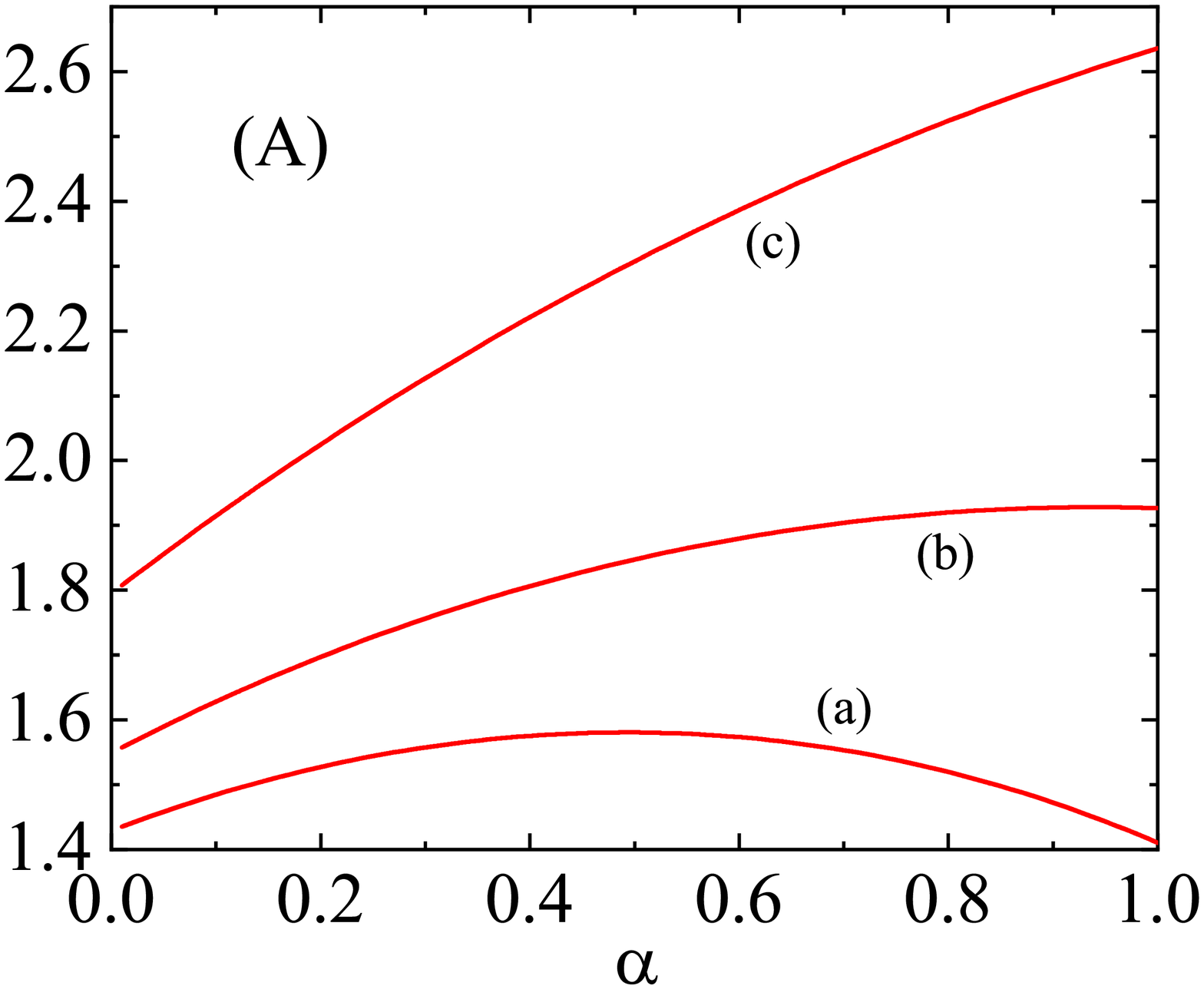}\hspace{5mm}
\includegraphics[width=0.35\textwidth]{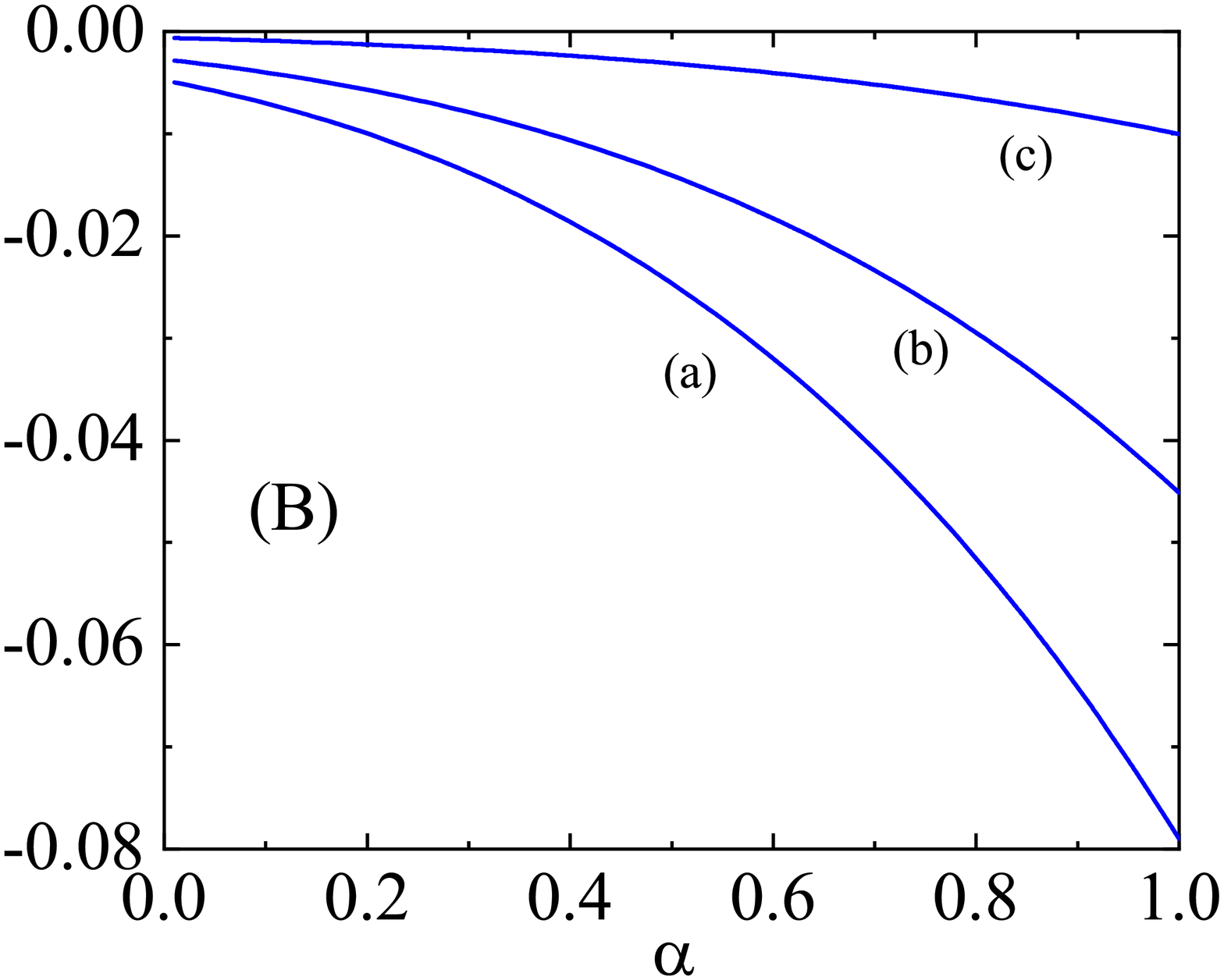}
\caption{Panel A: Dependence of the eigenvalues $\nu_{4|0}^{(11)}$ [defined by Eq.\ \eqref{b5}] (a), $\nu_{2|ij}^{(11)}$ [defined by Eq.\ \eqref{b7}] (b), and $\nu_{0|ijk\ell}^{(11)}$ [defined by Eq.\ \eqref{b9}] (c) on the (common) coefficient of restitution $\al$ for a three-dimensional binary mixture constituted by particles of the same mass density [$m_1/m_2=(\sigma_1/\sigma_2)^3$] with $x_1=\frac{1}{2}$ and $m_1/m_2=2$. Panel B: The same as panel A for the eigenvalues $\nu_{4|0}^{(12)}$ [defined by Eq.\ \eqref{b6}] (a), $\nu_{2|ij}^{(12)}$ [defined by Eq.\ \eqref{b8}] (b), and $\nu_{0|ijk\ell}^{(12)}$ [defined by Eq.\ \eqref{b10}] (c).
\label{fig3}}
\end{figure}

Apart from the partial temperatures, it is worthwhile analyzing the time evolution of the higher-degree velocity moments in the HCS. To get this equation, one takes velocity moments in both sides of Eqs.\ \eqref{2.1} and \eqref{2.1.0} and obtains the set of coupled equations
\begin{equation}
\label{2.5}
\partial_t M_{2p|\bar{q}}^{(1)}=J_{2p|\bar{q}}^{(11)}+J_{2p|\bar{q}}^{(12)},
\quad \partial_t M_{2p|\bar{q}}^{(2)}=J_{2p|\bar{q}}^{(22)}+J_{2p|\bar{q}}^{(21)}.
\end{equation}
In Eqs.\ \eqref{2.5}, we have introduced the short-hand notation $\bar{q}\equiv i_1i_2\ldots i_q$. To study the time evolution of the moments $\partial_t M_{2p|\bar{q}}^{(r)}$ it is convenient to introduce the \emph{scaled} moments
\begin{equation}
\label{2.4}
M_{2p|\bar{q}}^{*(r)}(t)\equiv n_r^{-1}[v_0(t)]^{-(2p+q)}M_{2p|\bar{q}}^{(r)}(t),
\end{equation}
where $v_0=\sqrt{2T(m_1+m_2)/m_1m_2}$ is a thermal velocity of the mixture. In accordance with Eq.\ \eqref{2.2}, one expects that after a transient regime the dimensionless moments $M_{2p|\bar{q}}^{*(r)}$ (scaled with the time-dependent thermal velocity $v_0(t)$) reach an asymptotic steady value.

The time evolution of the scaled moments $M_{2p|\bar{q}}^{*(r)}$ can be obtained when one takes into account the time evolution equation \eqref{2.5.1} for the temperature $T(t)$. In that case, from Eqs.\ (\ref{2.5.1}), \eqref{2.5}, and \eqref{2.4}, one simply gets
\begin{equation}
\label{2.6}
\partial_\tau  M_{2p|\bar{q}}^{*(1)}=J_{2p|\bar{q}}^{*(11)}+J_{2p|\bar{q}}^{*(12)}+\frac{2p+q}{2}\zeta^*
M_{2p|\bar{q}}^{*(1)},
\end{equation}
\begin{equation}
\label{2.6.0}
\partial_\tau  M_{2p|\bar{q}}^{*(2)}=J_{2p|\bar{q}}^{*(22)}+J_{2p|\bar{q}}^{*(21)}+\frac{2p+q}{2}\zeta^*
M_{2p|\bar{q}}^{*(2)},
\end{equation}
where $\zeta^*=\zeta/\nu_0'$,
\begin{equation}
\label{2.7}
J_{2p|\bar{q}}^{*(rs)}=\frac{J_{2p|\bar{q}}^{(rs)}}{\nu_0' n_rv_0^{2p+q}},
\end{equation}
and
\begin{equation}
\label{2.8}
\tau=\int_0^t\; ds\; \nu_0'(s).
\end{equation}
Since $\nu_0'(t)\propto T^{\beta}$ is an effective collision frequency, the parameter $\tau$ measures time as the number of (effective) collisions per particle. Here, for the sake of concreteness, we will consider Model B with $\beta=\frac{1}{2}$. In this case, as in previous works \cite{G03,GA05}, the effective collision frequency $\nu_0'(t)$ is
\beq
\label{nu0p}
\nu_0'(t)=\frac{\Omega_d}{4\sqrt{\pi}}n\sigma_{12}^{d-1}v_0(t).
\eeq
Needless to say, the results derived in this section are independent of the choice of $\nu_0'$; they apply for both Models A and B.

According to Eqs.\ \eqref{1.11}, \eqref{1.12}, and \eqref{1.13}--\eqref{1.18}, it is easy to see that the combination $J_{2p|\bar{q}}^{*(11)}+J_{2p|\bar{q}}^{*(12)}$ has the structure
\begin{equation}
\label{2.9}
J_{2p|\bar{q}}^{*(11)}+J_{2p|\bar{q}}^{*(12)}=-\nu_{2p|q}^{(11)}M_{2p|\bar{q}}^{*(1)}
-\nu_{2p|q}^{(12)}M_{2p|\bar{q}}^{*(2)}+\mathcal{C}_{2p|\bar{q}}^{(11)}
+\mathcal{C}_{2p|\bar{q}}^{(12)},
\end{equation}
where the terms $\mathcal{C}_{2p|\bar{q}}^{(rs)}$ are bilinear combinations of moments of degree less than $2p+q$. Since the first two terms on the right-hand side of Eq.\ \eqref{2.9} are linear, then the quantities $\nu_{2p|q}^{(11)}$ and $\nu_{2p|q}^{(12)}$ can be considered as the eigenvalues (or \emph{collisional rates}) of the linearized collision operators corresponding to the eigenfunctions $Y_{2p|\bar{q}}$. Their explicit forms for velocity moments of degree less than or equal to four are given in the Appendix \ref{appB}.

As an illustration, the dependence of the eigenvalues (collision rates) associated with the second, third, and fourth degree moments on the (common) coefficient of restitution $\al_{rs}\equiv \al$ is plotted in Figs.\ \ref{fig2} and \ref{fig3} for a binary mixture constituted by particles of the same mass density. Here, $d=3$, $x_1=\frac{1}{2}$, and $m_1/m_2=2$. While the eigenvalues  $\nu_{0|ij}^{(11)}$ and $\nu_{0|ijk}^{(11)}$ decrease with increasing inelasticity, the other two eigenvalues ($\nu_{2|0}^{(11)}$ and $\nu_{2|i}^{(11)}$) exhibit a non-monotonic dependence on $\al$. The eigenvalues of the second- and third-degree moments associated to cross-collisions ($\nu_{2|0}^{(12)}$, $\nu_{0|ij}^{(12)}$, $\nu_{2|i}^{(12)}$, and $\nu_{0|ijk}^{(12)}$) are negative and they increase with decreasing $\al$. A similar behavior can be found for the eigenvalues associated with the fourth-degree moments, as Fig.\ \ref{fig3} shows.  In general, we can conclude that the influence of inelasticity on those eigenvalues is in general important, specially in the case of the ones associated with the self-collisions (i.e., those of the form $\nu_{2p|q}^{(11)}$).

\subsection{Time evolution of the velocity moments}

Let us obtain the dependence of the (scaled) velocity moments $ M_{2p|\bar{q}}^{*(r)}$ on time. Thus, inserting the expression (\ref{2.9}) into Eq.\ (\ref{2.6}), in matrix form one finally gets
\begin{equation}
\label{2.10}
\left(\delta_{\sigma\sigma'}\partial_\tau+\mathcal{L}_{\sigma\sigma'}\right)
\mathcal{M}_{\sigma'}=\mathcal{C}_\sigma,
\end{equation}
where $\boldsymbol{\mathcal{M}}$ is the column matrix defined by the set
\begin{equation}
\label{2.12}
\left\{M_{2p|\bar{q}}^{*(1)}, M_{2p|\bar{q}}^{*(2)}\right\},
\end{equation}
and the square matrix $\boldsymbol{\mathcal{L}}$ is given by
\begin{equation}
\label{2.13}
\boldsymbol{\mathcal{L}}=\left(
\begin{array}{cc}
\omega_{2p|q}^{(11)}&\nu_{2p|q}^{(12)}\\
\nu_{2p|q}^{(21)}&\omega_{2p|q}^{(22)}
\end{array}
\right).
\end{equation}
In Eq.\ \eqref{2.13}, we have introduced the quantities
\begin{equation}
\label{2.14}
\omega_{2p|q}^{(11)}=\nu_{2p|q}^{(11)}-\frac{2p+q}{2}\zeta^*, \quad \omega_{2p|q}^{(22)}=\nu_{2p|q}^{(22)}-\frac{2p+q}{2}\zeta^*.
\end{equation}
The collision rates $\omega_{2p|q}^{(rs)}$ can be considered as \emph{shifted} collisional rates associated with the scaled moments $M_{2p|\bar{q}}^{*(r)}$.
Moreover, the column matrix $\boldsymbol{\mathcal{C}}$ is
\begin{equation}
\label{2.15}
\boldsymbol{\mathcal{C}}=\left(
\begin{array}{c}
{\mathcal C}_{2p|\bar{q}}^{(11)}+{\mathcal C}_{2p|\bar{q}}^{(12)}\\
{\mathcal C}_{2p|\bar{q}}^{(22)}+{\mathcal C}_{2p|\bar{q}}^{(21)}
\end{array}
\right).
\end{equation}
The solution of Eq.\ (\ref{2.10}) can be written as
\begin{equation}
\label{2.16}
\boldsymbol{\mathcal{M}}(\tau)=e^{-\boldsymbol{\mathcal{L}}\tau}\cdot \left[\boldsymbol{\mathcal {M}}(0)-
\boldsymbol{\mathcal{M}}(\infty)\right]+\boldsymbol{\mathcal{M}}(\infty),
\end{equation}
where the asymptotic steady value $\boldsymbol{\mathcal{M}}(\infty)$ is
\begin{equation}
\label{2.17}
\boldsymbol{\mathcal{M}}(\infty)=\boldsymbol{\mathcal{L}}^{-1}\cdot \boldsymbol{\mathcal{C}}.
\end{equation}

The long time behavior of $\mathcal{M}_\sigma$ $(\sigma=1,2)$ is governed by the smallest eigenvalue
$\ell_{2p|\bar{q}}^{\text{min}}$ of the matrix $\boldsymbol{\mathcal{L}}$. Given that the eigenvalues $\ell$ of the matrix $\boldsymbol{\mathcal{L}}$ obey the quadratic equation
\begin{equation}
\label{2.18}
\ell^2-(\omega_{2p|q}^{(11)}+\omega_{2p|q}^{(22)})\ell+(\omega_{2p|q}^{(11)}\omega_{2p|q}^{(22)}
-\nu_{2p|q}^{(12)}\nu_{2p|q}^{(21)})=0,
\end{equation}
the smallest eigenvalue $\ell_{2p|\bar{q}}^{\text{min}}$ is
\begin{equation}
\label{2.18.1}
\ell_{2p|\bar{q}}^{\text{min}}=\frac{\omega_{2p|q}^{(11)}+\omega_{2p|q}^{(22)}-\sqrt{\left(
\omega_{2p|q}^{(11)}+\omega_{2p|q}^{(22)}\right)^2-4\left(\omega_{2p|q}^{(11)}\omega_{2p|q}^{(22)}-
\nu_{2p|q}^{(12)}\nu_{2p|q}^{(21)}\right)}}{2}.
\end{equation}
If $\ell_{2p|\bar{q}}^{\text{min}}>0$, then all the scaled moments of degree $2p+q$ tend asymptotically to
finite values. Otherwise, for given values of the parameters of the mixture, if $\ell_{2p|\bar{q}}^{\text{min}}$ becomes negative for $\alpha_{rs}$ smaller than a certain critical value $\alpha_c$, then the moments of degree $2p+q$ exponentially grow in time for $\alpha_{rs}<\alpha_c$. The critical value $\alpha_c$ can be obtained from the condition
$\ell=0$ which implies
\begin{equation}
\label{2.19}
\omega_{2p|q}^{(11)}\omega_{2p|q}^{(22)}-\nu_{2p|q}^{(12)}\nu_{2p|q}^{(21)}=0.
\end{equation}

\section{Time behavior of the third and fourth degree moments in the HCS}
\label{sec5}

\begin{figure}
\includegraphics[width=0.4\textwidth]{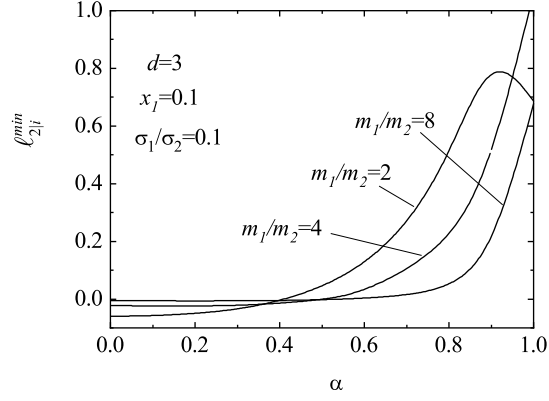}
\caption{Plot of the smallest eigenvalue $\ell^{\text{min}}_{2|i}$  associated with the long-time evolution of the (scaled) third degree (anisotropic) moments $\left\{M_{2|i}^{*(1)}, M_{2|i}^{*(2)}\right\}$ as a function of the (common) coefficient of restitution $\alpha$ for a three-dimensional system ($d=3$) with $x_1=0.1$, $\sigma_1/\sigma_2=0.1$ and three different values of the mass ratio $m_1/m_2$. The eigenvalue $\ell^{\text{min}}_{2|i}$ is defined by Eq.\ \eqref{2.18.1}.
\label{fig4}}
\end{figure}

\begin{figure}
\includegraphics[width=0.4\textwidth]{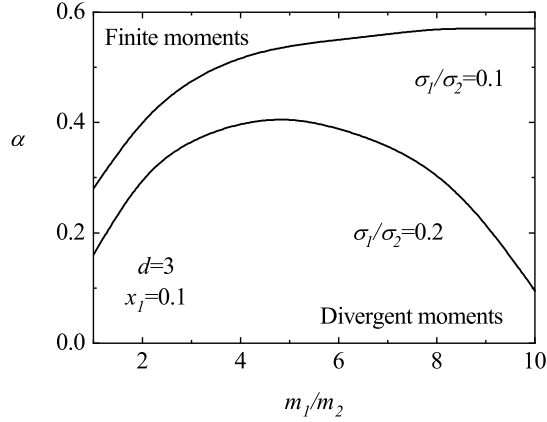}
\caption{Phase diagram in the ($\alpha, m_1/m_2$)--plane for the asymptotic long time behavior of the
third-degree moments for a three-dimensional system ($d=3$) with $x_1=0.1$ and two different values of
the size ratio $\sigma_1/\sigma_2$. The lines are obtained from the condition $\ell^{\text{min}}_{2|i}=0$.
\label{fig5}}
\end{figure}

The purpose of this section is to analyze the relaxation of the second, third and fourth degree moments to the HCS. However, a full analysis is difficult due to the many parameters involved in the problem:
$d, \alpha_{11}, \alpha_{12}, \alpha_{22}, m_1/m_2, x_1,$ and $\sigma_1/\sigma_2$. For the sake of concreteness,
henceforth we will consider the particular case $\alpha_{11}=\alpha_{12}=\alpha_{22}\equiv \alpha$. With respect to the second degree moments (those related to the elements of the pressure tensor), as expected our results show that these moments are convergent and tend asymptotically to well-defined values. In this context, it is important to remark that the reliability of the second degree moments of inelastic Maxwell mixtures have been tested against Monte Carlo simulations for IHS in the uniform shear flow problem \cite{G03}. The comparison between theory and simulations shows an excellent agreement between both approaches, even for quite strong dissipation and/or disparate values of the mass and diameter ratios. Let us analyze now the behavior of the third and fourth degree velocity moments.

\subsection{Third-degree moments}

In the case of the third degree moments, Eqs.\ (\ref{1.13}) and (\ref{1.14}) show that {\em all} the moments tend to zero for sufficiently long times provided the corresponding eigenvalues are negative. However, our analysis shows that while the moments $M_{2|ijk}^{*(r)}$ appear to be convergent (at least in the cases studied), the moments $M_{2|i}^{*(r)}$ (which are related to the heat flux) can be divergent since the corresponding eigenvalue $\ell^{\text{min}}_{2|i}$ can be negative. As an illustration, in Fig.\ \ref{fig4} we plot $\ell^{\text{min}}_{2|i}$ for $d=3$ with $x_1=\sigma_1/\sigma_2=0.1$ and three different values of the mass ratio ($m_1/m_2=2, 4$ and 8). It is apparent that, for given values of the parameters of the mixture, $\ell^{\text{min}}_{2|i}$ becomes negative for values of the coefficient of restitution smaller than a certain critical value $\alpha_c$. In particular, for the mixtures considered in Fig.\ \ref{fig4}, $\al_c\simeq 0.417, 0.516$, and 0.572 for $m_1/m_2=$2, 4, and 8, respectively. This means that, if $\alpha<\alpha_c$, the third degree moments $M_{2|i}^{(r)}$ exponentially grow in time. This singular behavior of the scaled third degree moments implies that the velocity distribution function $f_r(V)$ develops an algebraic high velocity tail in the long time limit of the form $f_r(V)\sim V^{-d-s}$. The exponent $s$ is quite sensitive to the values of the parameters of the mixture.  In particular, when the (scaled) moments $M_{2|i}^{*(r)}$ diverge in time for given values of the control parameters, then $s\leq 3$.

An study of the convergent/divergent regions of $M_{2|i}^{*(r)}$ is complex due to the parameter space of the binary system. However, an exhaustive analysis of the solutions to Eq.\ (\ref{2.19}) shows that the third degree moments $M_{2|i}^{(r)}$ appear usually to be convergent (i) when $\sigma_1>\sigma_2$ and $m_1>m_2$ , regardless the value of the mole fraction $x_1$, or (ii) when
$\sigma_1<\sigma_2$, $m_1>m_2$ but the mole fraction $x_1$ is not small enough (say, $x_1\gtrsim 0.2$). To illustrate these trends, Fig.\ \ref{fig5} shows a phase diagram associated with the singular behavior of the third-degree moments $M_{2|i}^{*(r)}$. Here, $d=3$, $x_1=0.1$ and two different values of the size ratio $\sigma_1/\sigma_2$ are considered. The curves $\alpha_c(m_1/m_2)$ split the parameter space into two regions: the region below the curve corresponds to states ($\alpha, m_1/m_2$) where the third-degree moments $M_{2|i}^{*(r)}$ diverge in time while the region above defines the states where those moments are convergent (and so they go to zero). It is apparent that the region of divergent moments grows as the size of the defect species decreases with respect to that of the excess component.
\begin{figure}
\includegraphics[width=0.4\textwidth]{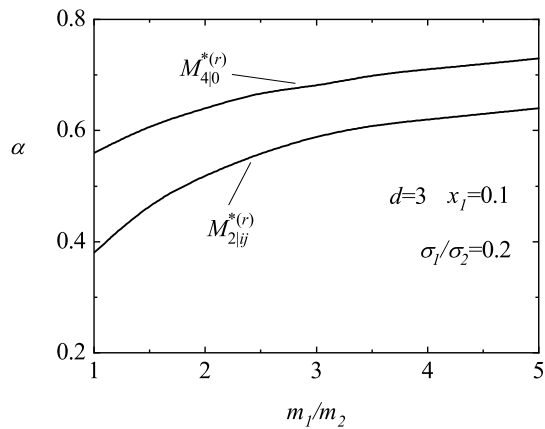}
\caption{Phase diagram in the ($\alpha, m_1/m_2$)--plane for the asymptotic long time behavior of the fourth-degree moments $M_{4|0}^{*(r)}$ and $M_{2|ij}^{*(r)}$ for a three-dimensional system ($d=3$) with $x_1=0.1$ and $\sigma_1/\sigma_2=0.2$. The lines are obtained from the condition $\ell^{\text{min}}_{2|i}=0$. \label{fig6}}
\end{figure}
\begin{figure}
\includegraphics[width=0.4\textwidth]{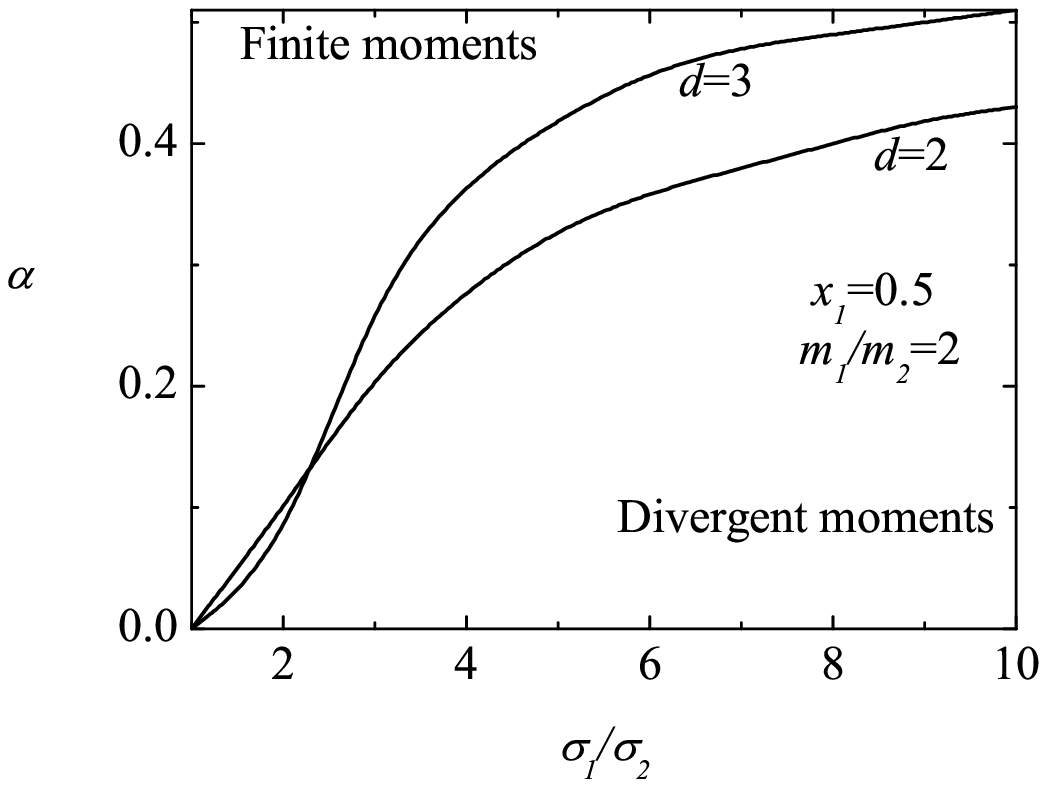}
\caption{Phase diagram in the ($\alpha, \sigma_1/\sigma_2$)--plane for the asymptotic long time behavior of the fourth-degree moments $M_{4|0}^{*(r)}$ for a three ($d=3$) and two ($d=2$) dimensional system with $x_1=0.5$ and $m_1/m_2=2$. The lines are obtained from the condition $\ell^{\text{min}}_{2|i}=0$. \label{fig7}}
\end{figure}
\begin{figure}
\includegraphics[width=0.4\textwidth]{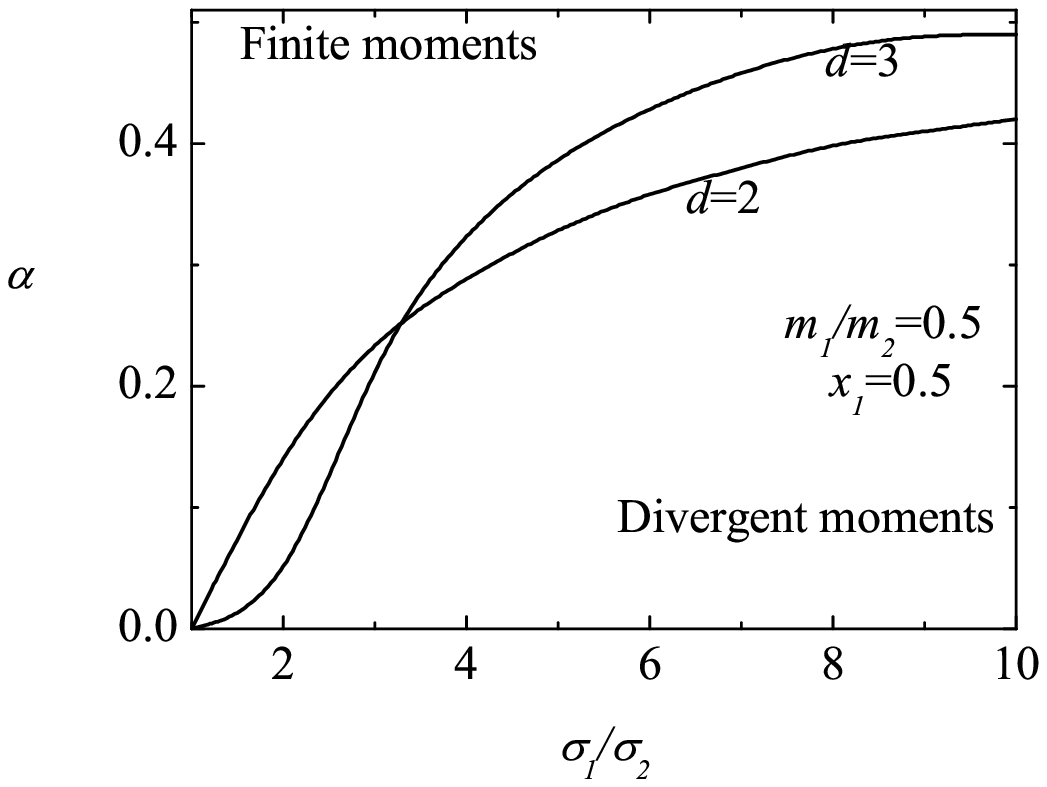}
\caption{Phase diagram in the ($\alpha, \sigma_1/\sigma_2$)--plane for the asymptotic long time behavior of the fourth-degree moments $M_{4|0}^{*(r)}$ for a three ($d=3$) and two ($d=2$) dimensional system with $x_1=0.5$ and $m_1/m_2=0.5$. The lines are obtained from the condition $\ell^{\text{min}}_{2|i}=0$. \label{fig8}}
\end{figure}

\subsection{Fourth-degree moments}

In the HCS state, $P_{r,ij}=p_r\delta_{ij}$ and so, the inhomogeneous terms ${\mathcal C}_{2p|q}^{(rs)}$ appearing in the right hand side of the expressions (\ref{1.15})--(\ref{1.18}) become
\begin{equation}
\label{2.20}
\mathcal {C}_{4|0}^{(rs)}=\frac{\omega_{rs}^*}{32}(1+\beta_{rs})^2
\frac{3\beta_{rs}^2-6\beta_{rs}+4d-1}{\theta_r\theta_s}, \quad \mathcal{C}_{2|ij}^{(rs)}=\mathcal{C}_{0|ijk\ell}^{(rs)}=0.
\end{equation}
Consequently, if the eigenvalues of the matrix
$\boldsymbol{\mathcal{L}}$ corresponding to the fourth-degree moments are positive, Eq. (\ref{2.20}) shows that all the moments, except $M_{4|0}^{*(r)}$ tend to zero for $\tau\to \infty$. The asymptotic expression of $M_{4|0}^{*(r)}$ is
\begin{equation}
\label{2.21}
M_{4|0}^{*(1)}=\frac{\omega_{4|0}^{(22)}(\mathcal{C}_{4|0}^{(11)}+\mathcal{C}_{4|0}^{(12)})-\nu_{4|0}^{(12)}
(\mathcal{C}_{4|0}^{(22)}+\mathcal{C}_{4|0}^{(21)})}
{\omega_{4|0}^{(11)}\omega_{4|0}^{(22)}-\nu_{4|0}^{(12)}\nu_{4|0}^{(21)}}.
\end{equation}
The expression for $M_{4|0}^{*(2)}$ can be easily obtained from (\ref{2.21}) by just making the changes $1\leftrightarrow 2$. Note that when the condition (\ref{2.19}) applies, the fourth-degree moments $M_{4|0}^{*(r)}$ tend to infinite as expected.

An analysis on the possible divergence of the fourth-degree moments clearly shows that these moments can be also divergent in some regions of the parameter space of the system, specially the (isotropic) moment $M_{4|0}^{*(r)}$. Figure \ref{fig6} illustrates this fact for $d=3$, $x_1=0.1$ and $\sigma_1/\sigma_2=0.2$. In this case, while the moments $M_{0|ijk\ell}^{*(r)}$ are convergent (and tend to zero for long times), the moments $M_{4|0}^{*(r)}$ and $M_{2|ij}^{*(r)}$ can be divergent (regions below the curves). It is also apparent that for a given value of the mass ratio, the critical value of the coefficient of restitution of $M_{4|0}^{*(r)}$ is larger than the one found for $M_{2|ij}^{*(r)}$ so that the divergent region of the former is bigger than the latter one. Figures \ref{fig7} and \ref{fig8} complement the results shown in Fig.\ \ref{fig6}. We plot the phase diagram of $M_{4|0}^{*(r)}$ in the ($\alpha, \sigma_1/\sigma_2$)-plane for spheres ($d=3$) and disks ($d=2$). We observe that the influence of the mass ratio is not quite significant on the form of the phase diagram. Moreover, while the divergent region seems to be more important for disks than for hard spheres when the size ratio is small (but larger than one), the opposite happens as the size ratio increases.

\section{USF. Tracer limit}
\label{sec6}

As a second application, we study in this section the USF problem. This state is
macroscopically characterized by constant densities $n_r$, a uniform temperature $T$, and a
linear velocity profile
\begin{equation}
\label{6.1}
\mathbf{U}(y)=\mathbf{U}_1(y)=\mathbf{U}_2(y)=ay \widehat{\mathbf {x}},
\end{equation}
where $a$ is the constant shear rate. This linear velocity profile assumes no boundary
layer near the walls and is generated by the Lees-Edwards boundary conditions \cite{LE72}, which are simply periodic boundary conditions in the local Lagrange frame moving with the flow velocity. Since $n_r$ and $T$ are uniform, then the mass and heat fluxes vanish and the transport of momentum (measured by the pressure
tensor) is the relevant phenomenon. At a microscopic level, the USF is characterized by
a velocity distribution function that becomes \emph{uniform} in the local Lagrangian
frame, i.e., $f_{s}({\bf r},{\bf v},t)=f_{s}({\bf V},t)$. In that case, the Boltzmann equation for the binary mixture is given by the set of coupled kinetic equations
\begin{equation}
\label{6.2}
\frac{\partial}{\partial t}f_1-aV_y\frac{\partial}{\partial V_x}f_1=J_{11}[f_1,f_1]+J_{12}[f_1,f_2],
\end{equation}
\begin{equation}
\label{6.3}
\frac{\partial}{\partial t}f_2-aV_y\frac{\partial}{\partial V_x}f_2=J_{22}[f_2,f_2]+J_{21}[f_2,f_1].
\end{equation}
Equations \eqref{6.2} and \eqref{6.3} are invariant under the changes $\left(V_x,V_y\right)\to \left(-V_x,-V_y\right)$ and $V_j\to -V_j$ ($j\neq x,y$).

The relevant macroscopic balance equation in the USF state is the
balance equation for the temperature $T=(1/d n)(m_1 M_{2|0}^{(1)}+m_2 M_{2|0}^{(2)})$. This equation can be easily obtained from Eqs.\ \eqref{6.2} and \eqref{6.3}. In dimensionless form, it can be written as
\begin{equation}
\label{6.4}
\nu_0^{-1}\frac{\partial}{\partial t}\ln T=-\zeta^*-\frac{2a^*}{d} P_{xy}^*,
\end{equation}
where $\nu_0\propto n T^{\beta}$, $\zeta^*=\zeta/\nu_0$, $a^*=a/\nu_0$, $P_{xy}^*=P_{xy}/p$, $p=nT=p_1+p_2$ being the
hydrostatic pressure. Equation (\ref{6.4}) shows that the temperature changes in time
due to the competition of two opposite mechanisms: on the one hand, viscous heating
($-a^*P_{xy}^*>0$) and, on the other hand, energy dissipation in collisions ($-\zeta^*<0$). The
{\em reduced} shear rate $a^*$ is the nonequilibrium relevant parameter of the USF
problem since it measures the departure of the system from the HCS (vanishing shear rate). It is apparent that, except for Model A ($\beta=0$), the (reduced) shear rate $a^*(t)\propto
T(t)^{-\beta}$ is a function of time. Therefore, for $\beta\neq 0$ (model B), after a transient regime a steady state
is achieved in the long time limit when both viscous heating and collisional cooling
cancel each other and the mixture autonomously seeks the temperature at which the above
balance occurs. In this steady state, the reduced steady shear rate $a_{\text{st}}^*$ and the coefficients of
restitution $\al_{rs}$ are not independent parameters since they are related through the \emph{steady}
state condition
\beq
\label{6.5}
a_{\text{st}}^*P_{\text{st},xy}^*=-\frac{d}{2}\zeta_{\text{st}}^*,
\eeq
where the subindex $\text{st}$ means that the quantities are evaluated in the steady state. However, when $\beta=0$ (model A), $\nu_0\equiv \text{const}.$,  $\partial_t a^*=0$ and so, the reduced shear rate
remains in its initial value regardless of the values of the coefficients of
restitution $\alpha_{rs}$. As a consequence, there is no steady state (unless $a^*$
takes the specific value given by the condition (\ref{6.5})) and $a^*$ and
$\alpha_{rs}$ are \emph{independent} parameters in the USF problem. This is one of the main advantages of using Model A instead of Model B in the USF problem.

Before going ahead, it is convenient to write the form of $\omega_{rs}$ for arbitrary values of $\beta$. Here, although we will mainly consider model A, as in previous works on IMM \cite{GT10,GT15} we will keep the same form for $\omega_{rs}$ as in model B with $\beta=\frac{1}{2}$. Thus, $\omega_{rs}$ can be written as \cite{GT10,GT15}
\beq
\label{w}
\omega_{rs}=x_s\left(\frac{\sigma_{rs}}{\sigma_{12}}\right)^{d-1}
\left(\frac{\theta_r+\theta_s}{\theta_r\theta_s}\right)^{1/2}\nu_0, \quad \nu_0=A(\beta) n T^\beta,
\eeq
where the value of the quantity $A(\beta)$ is irrelevant for our calculations. In Eq.\ \eqref{w},
\beq
\label{6.6}
\theta_r=\frac{m_r T}{T_r}\sum_{s=1}^2 m_s^{-1}.
\eeq

As in the case of the HCS, to determine the hierarchy of moment equations in the USF we multiply both sides of Eqs.\ \eqref{6.1} and \eqref{6.2} by $Y_{2p|\overline{q}}^{(r)}(\mathbf{V})$ and integrates over $\mathbf{V}$. The result is
\beq
\label{6.7}
\partial_t M_{2p|\overline{q}}^{(1)}+a N_{2p|\overline{q}}^{(1)}=J_{2p|\overline{q}}^{(11)}+J_{2p|\overline{q}}^{(12)},
\eeq
\beq
\label{6.8}
\partial_t M_{2p|\overline{q}}^{(2)}+a N_{2p|\overline{q}}^{(2)}=J_{2p|\overline{q}}^{(22)}+J_{2p|\overline{q}}^{(21)}.
\eeq
Here, we have called
\beq
\label{6.9}
N_{2p|\overline{q}}^{(r)}=\int d\mathbf{V} f_r(\mathbf{V}) V_y \frac{\partial}{\partial V_x}Y_{2p|\overline{q}}^{(r)}(\mathbf{V}).
\eeq
In particular,
\beq
\label{6.10}
N_{2|0}^{(r)}=2 M_{0|xy}^{(r)}, \quad N_{0|yy}^{(r)}=-\frac{2}{d} M_{0|xy}^{(r)}, \quad N_{0|xy}^{(r)}=M_{0|yy}^{(r)}+\frac{1}{d} M_{2|0}^{(r)}.
\eeq
Since $V_y \partial_{V_x}Y_{2p|\overline{q}}^{(r)}(\mathbf{V})$ is a polynomial of degree $2p+q$, then the quantity $N_{2p|\overline{q}}^{(r)}$ can be expressed as linear combinations of moments of degree $2p+q$. This means that the hierarchy of Eqs.\ \eqref{6.7} and \eqref{6.8} can be exactly solved in a recursive way. This contrasts with the set of coupled equations for the moments in the HCS where a general solution for them can be formally written.

Due to the technical difficulties involved in the solution of Eqs.\ \eqref{6.7} and \eqref{6.8}  for a general binary mixture, we consider here the limit case where the concentration of one of the species (let's say, species 1) is negligible ($x_1\to 0$). This is the so-called \emph{tracer} limit. In this situation, one can assume that the state of the excess component $2$ is not perturbed by the presence of the tracer particles and so, Eq.\ \eqref{6.8} reduces to
\beq
\label{6.11}
\partial_t M_{2p|\overline{q}}^{(2)}+a N_{2p|\overline{q}}^{(2)}=J_{2p|\overline{q}}^{(22)}.
\eeq
On the other hand, one can also neglect the collisions among tracer particles themselves in Eq.\ \eqref{6.7} and so, this equations reads
\beq
\label{6.12}
\partial_t M_{2p|\overline{q}}^{(1)}+a N_{2p|\overline{q}}^{(1)}=J_{2p|\overline{q}}^{(12)}.
\eeq

For the sake of convenience, let us introduce the scaled moments
\begin{equation}
\label{6.13}
M_{2p|\bar{q}}^{*(r)}(t)\equiv n_2^{-1}[v_{02}t)]^{-(2p+q)}M_{2p|\bar{q}}^{(r)}(t),
\end{equation}
where $v_{02}(t)=\sqrt{2T_2(t)/m_2}$ is the thermal velocity of the excess species. Note that in the tracer limit $n\simeq n_2$ and $T(t)\simeq T_2(t)$. The evolution equations for the scaled moments $M_{2p|\bar{q}}^{*(r)}(t)$ can be obtained from Eqs.\ \eqref{6.11} and \eqref{6.12} when one takes into account the balance equation \eqref{6.4} for the temperature $T(t)$. They are given by
\beq
\label{6.14}
\partial_\tau M_{2p|\overline{q}}^{*(2)}-\left(p+\frac{q}{2}\right)\left(\zeta_0^*+\frac{4a^*}{d}M_{0|xy}^{*(2)}\right)
M_{2p|\overline{q}}^{*(1)}+a^* N_{2p|\overline{q}}^{*(2)}=J_{2p|\overline{q}}^{*(22)},
\eeq
\beq
\label{6.15}
\partial_\tau M_{2p|\overline{q}}^{*(1)}-\left(p+\frac{q}{2}\right)\left(\zeta_0^*+\frac{4a^*}{d}M_{0|xy}^{*(2)}\right)
M_{2p|\overline{q}}^{*(1)}+a^* N_{2p|\overline{q}}^{*(1)}=J_{2p|\overline{q}}^{*(12)},
\eeq
where
\beq
\label{6.16}
N_{2p|\bar{q}}^{*(r)}\equiv n_2^{-1}[v_{02}]^{-(2p+q)}N_{2p|\bar{q}}^{(r)}, \quad J_{2p|\overline{q}}^{*(rs)}=\frac{1}{\nu_0 n_2 v_{02}^{2p+q}}J_{2p|\overline{q}}^{(rs)},
\eeq
and
\beq
\label{6.17}
\zeta_0^*=\frac{\zeta}{\nu_0}=\frac{1-\al_{22}^2}{2d}\omega_{22}^*, \quad \omega_{22}^*=\frac{\omega_{22}}{\nu_0}=\left(\frac{\sigma_2}{\sigma_{12}}\right)^{d-1}\sqrt{2\mu_{12}}.
\eeq
Upon writing Eqs.\ \eqref{6.14} and \eqref{6.15}, use has been made of the identity $P_{xy}^*=2M_{0|xy}^{*(2)}$ and the definition \eqref{2.8} of $\tau$ with the replacement $\nu_0'\to \nu_0$.

As expected, the evolution equations \eqref{6.15} and \eqref{6.16} involve the (reduced) shear stress $M_{0|xy}^{*(2)}$ (second-degree moment). Thus, to determine the time evolution of the high-degree moments in the USF, one has to get first the second-degree moments. These moments are the most relevant ones from a rheological point of view.

\subsection{Second-degree moments of the excess species. Model A}

In the case of the excess component, the set of coupled equations for the moments $M_{0|xy}^{*(2)}$ and $M_{0|yy}^{*(2)}$ can be easily obtained from Eq.\ \eqref{6.14}:
\beq
\label{6.18}
\partial_\tau M_{0|xy}^{*(2)}+a^*\left(M_{0|yy}^{*(2)}+\frac{1}{2}\right)
+\left(\omega_{0|2}^{(22)}-\frac{4a^*}{d}M_{0|xy}^{*(2)}\right)M_{0|xy}^{*(2)}=0,
\eeq
\beq
\label{6.19}
\partial_\tau M_{0|yy}^{*(2)}-\frac{2}{d}a^*M_{0|xy}^{*(2)}
+\left(\omega_{0|2}^{(22)}-\frac{4a^*}{d}M_{0|xy}^{*(2)}\right)M_{0|yy}^{*(2)}=0,
\eeq
where in the tracer limit
\beq
\label{6.20}
\omega_{0|2}^{(22)}=\frac{(1+\al_{22})^2}{2(d+2)}\omega_{22}^*.
\eeq

In the hydrodynamic regime (which holds for times longer than the mean free time), the dependence of the (scaled) moments $M_{0|xy}^{*(2)}$ and $M_{0|yy}^{*(2)}$ on the dimensionless time $\tau$ is via the time-dependence of the reduced shear rate $a^*(\tau)$. Therefore, in Model A, $\partial_\tau M_{0|ij}^{*(2)}=0$ (since $a^*\text{const}.$) and  the scaled moments $M_{0|yy}^{*(2)}$ and $M_{0|xy}^{*(2)}$ achieve stationary values which are nonlinear functions of $a^*$ and $\al_{22}$. Their expressions are \cite{SG07}
\beq
\label{6.21}
M_{0|yy}^{*(2)}=-\frac{\Lambda(\widetilde{a})}{1+2\Lambda(\widetilde{a})}, \quad M_{0|xy}^{*(2)}=-\frac{1}{2}\frac{\widetilde{a}}{\left[1+2\Lambda(\widetilde{a})\right]^2},
\eeq
where $\Lambda(\widetilde{a})$ is the real root of the cubic equation
\beq
\label{6.22}
\Lambda(1+2\Lambda)^2=\frac{\widetilde{a}^2}{d},
\eeq
namely,
\beq
\label{6.23}
\Lambda(\widetilde{a})=\frac{2}{3}\sinh^2\left[\frac{1}{6}\cosh^{-1}\left(1+\frac{27}{d}\widetilde{a}^2\right)\right].
\eeq
Here,
\beq
\label{6.24}
\widetilde{a}=\frac{a^*}{\omega_{0|2}^{(22)}}=\frac{2(d+2)}{(1+\al_{22})^2}\frac{a^*}{\omega_{22}^*}.
\eeq

\section{Second and third degree moments of the tracer species. Model A}
\label{sec7}

In this section, we study the time evolution of the second and third-degree moments of the tracer species within the context of Model A. In particular, to obtain the time evolution of the scaled second-degree moments $M_{0|ij}^{*(1)}$ of the tracer species, we assume that the scaled moments $M_{0|ij}^{*(2)}$ have reached their stationary values. Therefore, from Eq.\ \eqref{6.15} one gets the set of coupled equations
\beq
\label{6.25}
\partial_\tau M_{0|xy}^{*(1)}+a^*\left(M_{0|yy}^{*(1)}+\frac{x_1 \gamma}{2\mu}\right)
+\left(\omega_{0|2}^{(12)}+2\omega_{0|2}^{(22)}\Lambda\right)M_{0|xy}^{*(1)}=x_1\frac{(1+\beta_{12})^2}{2d(d+2)}\omega_{12}^*
M_{0|xy}^{*(2)},
\eeq
\beq
\label{6.26}
\partial_\tau M_{0|yy}^{*(1)}-\frac{2}{d}a^*M_{0|xy}^{*(1)}
+\left(\omega_{0|2}^{(12)}+2\omega_{0|2}^{(22)}\Lambda\right)M_{0|yy}^{*(1)}=x_1\frac{(1+\beta_{12})^2}{2d(d+2)}\omega_{12}^*
M_{0|yy}^{*(2)},
\eeq
\beq
\label{6.27}
\partial_\tau M_{0|xx}^{*(1)}+\frac{2(d-1)}{d}a^*M_{0|xy}^{*(1)}
+\left(\omega_{0|2}^{(12)}+2\omega_{0|2}^{(22)}\Lambda\right)M_{0|xx}^{*(1)}=x_1\frac{(1+\beta_{12})^2}{2d(d+2)}\omega_{12}^*
M_{0|xx}^{*(2)},
\eeq
where $\gamma=T_1/T_2$ is the temperature ratio, $\mu=m_1/m_2$ is the mass ratio,
\beq
\label{6.28}
\omega_{0|2}^{(12)}=\frac{\omega_{12}^*}{2d(d+2)}(1+\beta_{12})(2d+3-\beta_{12})-\zeta^*,
\eeq
and
\beq
\label{6.29}
\omega_{12}^*=\frac{\omega_{12}}{\nu_0}=\sqrt{\mu_{12}+\mu_{21}\gamma},
\eeq
Upon writing Eqs.\ \eqref{6.25}--\eqref{6.27}, we have taken into account the relationship $(4a^*/d)M_{0|xy}^{*(2)}=-2\omega_{0|2}^{(22)}\Lambda$. Note that the moments associated with the tracer species are proportional to $x_1$. For this reason, the right hand side of Eqs.\ \eqref{6.25}--\eqref{6.27} are proportional to $x_1$.

\begin{figure}
\includegraphics[width=0.4\textwidth]{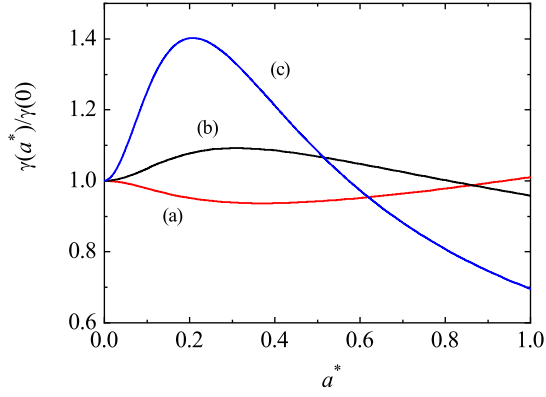}
\caption{Shear-rate dependence of the ratio $\gamma(a^*)/\gamma(0)$ for a three-dimensional mixture with $\sigma_1/\sigma_2=1$, $\al_{22}=\al_{12}=0.8$, and three different values of the mass ratio: $m_1/m_2=0.5$ (a), $m_1/m_2=2$ (b), and $m_1/m_2=5$ (c). The temperature ratio is obtained by numerically solving Eq.\ \eqref{6.35}.
\label{fig9}}
\end{figure}

For long times, in the case of Model A, $\partial_\tau M_{0|ij}^{*(1)}\to 0$ and so, the solution to Eqs.\ \eqref{6.25} and \eqref{6.26} can be written as
\beq
\label{6.30}
M_{0|yy}^{*(1)}=\frac{x_1}{\Delta}\Bigg[\frac{2a^*}{d}\Big(B M_{0|xy}^{*(2)}-\frac{a^*\gamma}{2\mu}\Big)+B C M_{0|yy}^{*(2)}\Bigg],
\eeq
\beq
\label{6.31}
M_{0|xy}^{*(1)}=\frac{x_1}{\Delta}\Bigg[C\Big(B M_{0|xy}^{*(2)}-\frac{a^*\gamma}{2\mu}\Big)-a^*B  M_{0|yy}^{*(2)}\Bigg],
\eeq
where
\beq
\label{6.32}
B\equiv \frac{(1+\beta_{12})^2}{2d(d+2)}\omega_{12}^*, \quad C\equiv \omega_{0|2}^{(12)}+2\omega_{0|2}^{(22)}\Lambda, \quad
\Delta \equiv C^2+\frac{2a^{*2}}{d}.
\eeq
In terms of $M_{0|xy}^{*(1)}$, the expression of $M_{0|xx}^{*(1)}$ is
\beq
\label{6.33}
M_{0|xx}^{*(1)}=\frac{x_1}{C}\Bigg[B M_{0|xx}^{*(2)}-\frac{2(d-1)}{d}x_1^{-1}a^*M_{0|xy}^{*(1)}\Bigg].
\eeq
As expected, from Eqs.\ \eqref{6.30}, \eqref{6.31}, and \eqref{6.32}, it is straightforward to verify the constraint
\beq
\label{6.34}
M_{0|xx}^{*(1)}+(d-1)M_{0|yy}^{*(1)}=0.
\eeq
Equations \eqref{6.30}, \eqref{6.31}, and \eqref{6.32} are consistent with the results obtained in the Appendix C of Ref.\ \cite{GT15}.

\begin{figure}
\includegraphics[width=0.4\textwidth]{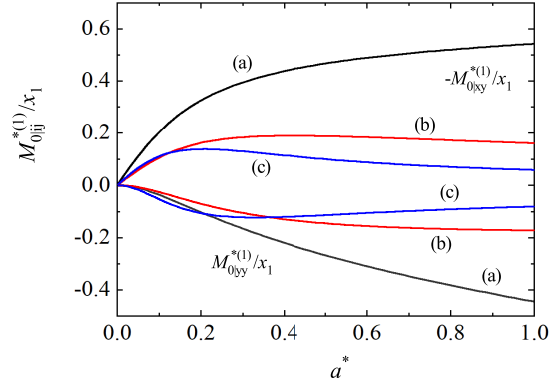}
\caption{Shear-rate dependence of the scaled moments of the tracer species $-x_1^{-1}M_{0|xy}^{*(1)}$ [defined by Eq.\ \eqref{6.30}] and $x_1^{-1} M_{0|yy}^{*(1)}$ [defined by Eq.\ \eqref{6.31}] for a three-dimensional mixture with $\sigma_1/\sigma_2=1$, $\al_{22}=\al_{12}=0.8$, and three different values of the mass ratio: $m_1/m_2=0.5$ (a), $m_1/m_2=2$ (b), and $m_1/m_2=5$ (c).
\label{fig10}}
\end{figure}

It still remains to determine the temperature ratio $\gamma$. This quantity can be obtained by combining the balance equations for the temperatures $T_2$ and $T_1$. In the case of Model A, $\gamma$ is determined by numerically solving the equation
\beq
\label{6.35}
\gamma \zeta_1^*+\frac{2a^*}{d}x_1^{-1} P_{1,xy}^*=\gamma\Big(\zeta^*+\frac{2a^*}{d}P_{2,xy}^*\Big),
\eeq
where $P_{2,xy}^*=2 M_{0|xy}^{*(2)}$, $P_{1,xy}^*=2\mu M_{0|xy}^{*(1)}$, and
\beq
\label{6.36}
\zeta_1^*=\frac{\omega_{12}^*}{4d}(1+\beta_{12})\left[3-\beta_{12}-(1+\beta_{12})\frac{\mu}{\gamma}\right].
\eeq
For mechanically equivalent particles, $x_1^{-1}M_{0|ij}^{*(1)}= M_{0|ij}^{*(2)}$ and the condition \eqref{6.35} yields $\gamma=1$ for any value of both the shear rate and the coefficients of restitution. This is the expected result. Moreover, when $a^*=0$ and $\al_{rs}\neq 1$, one recovers the results obtained in the tracer limit of the HCS. To illustrate the shear-rate dependence of the temperature ratio, we plot in Fig.\ \ref{fig9} the ratio $\gamma(a^*)/\gamma(0)$ versus the (reduced) shear rate $a^*$ for $d=3$, $\sigma_1/\sigma_2=1$, $\al_{22}=\al_{12}=0.8$, and three different values of the mass ratio. Here, $\gamma(0)$ is the value of the temperature ratio in the HCS. We observe first that the influence of $a^*$ on $\gamma$ is significant since the ratio $\gamma(a^*)/\gamma(0)$ clearly differs from 1. In addition, in contrast to the results obtained in the HCS [see Fig.\ \ref{fig1}], the temperature ratio $\gamma$ exhibits a non-monotonic dependence on $a^*$ regardless of the mass ratio considered. To complement Fig.\ \ref{fig9}, Fig.\ \ref{fig10} shows the shear-rate dependence of the scaled moments $x_1^{-1}M_{0|xy}^{*(1)}$ and $x_1^{-1} M_{0|yy}^{*(1)}$ for the same systems as that of Fig.\ \ref{fig9}. While the first moment is related with the tracer contribution to the shear stress, the second moment is a measure of the normal stress differences. It is quite apparent that the non-Newtonian effects on tracer species increase as increasing the shear rate, as expected. In addition, the departure from equilibrium becomes more significant as the tracer species is lighter than the excess species.

\subsection{Third-degree moments. Model A}

We consider now the time evolution of the (scaled) third-degree moments in the context of Model A. Let us assume first that the scaled second-degree moments have achieved their stationary values given by Eqs.\ \eqref{6.21} for the excess species and Eqs.\ \eqref{6.30}, \eqref{6.31}, and \eqref{6.33} for the tracer species. Moreover, as shown in Ref.\ \cite{SG07}, all the scaled third-degree moments of the excess species vanish for long times in the USF. Here, as in the analysis of the second-degree moments of the tracer species, we also assume that the scaled moments $M_{2|i}^{*(2)}$ and $M_{0|ijk}^{*(2)}$ have reached their steady values (and so, they vanish). In what follows, for the sake of simplicity, we will particularize to a two-dimensional system ($d=2$).

\begin{figure}
\includegraphics[width=0.4\textwidth]{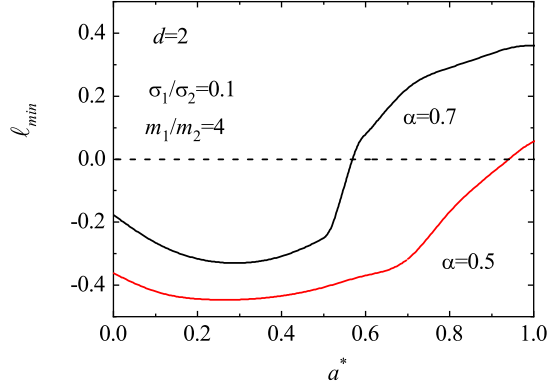}
\caption{Plot of the smallest eigenvalue $\ell_\text{min}$ associated with the time-evolution of the third-degree moments as a function of the reduced shear rate $a^*$ for $d=2$, $\sigma_1/\sigma_2=0.1$, $m_1/m_2=4$, and two values of the (common) coefficient of restitution: $\al=0.7$ and $\al=0.5$. The eigenvalue $\ell_\text{min}$ refers to the eigenvalue of Eq.\ \eqref{6.41} with the smallest real part.
\label{fig11}}
\end{figure}
\begin{figure}
\includegraphics[width=0.4\textwidth]{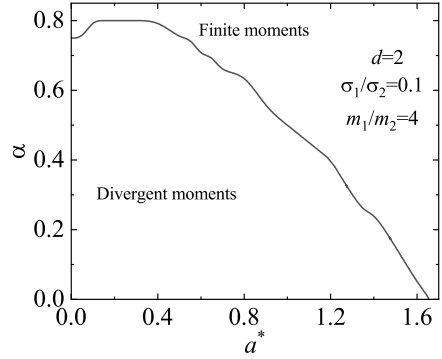}
\caption{Phase diagram in the $\left(\al, a^* \right)$-plane for the asymptotic long time behavior of the third-degree moments of the tracer species in a binary mixture under USF. We consider here a two-dimensional system ($d=2$) with $\sigma_1/\sigma_2=0.1$ and $m_1/m_2=4$. The region below the line corresponds to states where the scaled third-degree moments diverge in time while the region above the line refers to states where those moments vanish. The line is obtained from the condition $\ell_\text{min}=0$.
 \label{fig12}}
\end{figure}
In a two-dimensional mixture, there are 4 independent third-degree moments for the tracer species. Here we take the scaled
moments
\beq
\label{6.37}
\left\{M_{2|x}^{*(1)}, M_{2|y}^{*(1)}, M_{0|xxy}^{*(1)}, M_{0|xyy}^{*(1)}\right\}.
\eeq
After some algebra, the time evolution of the moments \eqref{6.37} is given by
\beq
\label{6.38}
\left(
\begin{array}{cccc}
\partial_\tau+W_{2|1}&\frac{3}{2}a^*&2a^*&0\\
\frac{a^*}{2}&\partial_\tau+W_{2|1}&0&2a^*\\
\frac{3}{8}a^*&0&\partial_\tau+W_{0|3}&\frac{3}{2}a^*\\
0&\frac{3}{8}a^*&-\frac{3}{2}a^*&\partial_\tau+W_{0|3}
\end{array}
\right)
\left(
\begin{array}{c}
M_{2|x}^{*(1)}\\
M_{2|y}^{*(1)}\\
M_{0|xxy}^{*(1)}\\
M_{0|xyy}^{*(1)}
\end{array}
\right)=
\left(
\begin{array}{c}
0\\
0\\
0\\
0
\end{array}
\right),
\eeq
where
\beq
\label{6.42}
W_{2|1}\equiv \omega_{2|1}^{(12)}+3\omega_{0|2}^{(22)}\Lambda, \quad W_{0|3}\equiv \omega_{0|3}^{(12)}+3\omega_{0|2}^{(22)}\Lambda,
\eeq
\beq
\label{6.39}
\omega_{2|1}^{(12)}=\frac{\omega_{12}^*}{8d(d+2)}(1+\beta_{12})[3\beta_{12}^2-2(d+5)\beta_{12}+10d+11]-\frac{3}{2}\zeta^*,
\eeq
and
\beq
\label{6.40}
\omega_{0|3}^{(12)}=3\frac{\omega_{12}^*}{4d(d+2)(d+4)}(1+\beta_{12})[\beta_{12}^2-2(d+3)
\beta_{12}+2d^2+10d+9]-\frac{3}{2}\zeta^*.
\eeq

In the absence of shear rate ($a^*=0$), the eigenvalues associated with the moments $\left(M_{2|x}^{*(1)}, M_{2|y}^{*(1)}\right)$ and $\left(M_{0|xxy}^{*(1)}, M_{0|xyy}^{*(1)}\right)$ are $\omega_{2|1}^{(12)}$ and $\omega_{0|3}^{(12)}$, respectively. This result agrees with the ones obtained in the HCS in the tracer limit when one assumes that the (scaled) third-degree moments of the excess component vanish. Thus, in the HCS, the moments $\left(M_{2|x}^{*(1)}, M_{2|y}^{*(1)}\right)$ and $\left(M_{0|xxy}^{*(1)}, M_{0|xyy}^{*(1)}\right)$ are divergent if $\omega_{2|1}^{(12)}$ and $\omega_{0|3}^{(12)}$ are negative, respectively.

When $a^*\neq 0$, the eigenvalues $\ell$ associated with the time behavior of the third-degree moments \eqref{6.37} are the roots of the characteristic quartic equation
\beq
\label{6.41}
\left(W_{0|3}-\ell\right)^2\left(W_{2|1}-\ell\right)^2=\frac{3}{4}a^{*2}\Big(W_{0|3}-W_{2|1}\Big)\Big(W_{0|3}+3W_{2|1}-4\ell\Big).
\eeq
The long time behavior of the moments  \eqref{6.37} is governed by the eigenvalue $\ell_\text{min}$ with the smallest real part. If $\ell_\text{min}$ becomes negative then the third-degree moments of the tracer species can be divergent.

As expected, an analysis of the solutions of the quartic equation \eqref{6.41} shows that $\ell_\text{min}$ may be negative, specially when the diameter of the tracer species is smaller than that of the excess species. Moreover, surprisingly, in most of the cases studied we have found that the main effect of shear rate on $\ell_\text{min}$ is to reduce its magnitude so that, it becomes positive for shear rates larger than a certain critical value. As an illustration, Fig.\ \ref{fig11} shows the dependence of $\ell_\text{min}$ on $a^*$ for $d=2$, $\sigma_1/\sigma_2=0.1$, $m_1/m_2=4$, and two values of the (common) coefficient of restitution. We observe that $\ell_\text{min}$ is a non-monotonic function of the shear rate; it becomes positive for sufficiently large values of $a^*$. To complement Fig.\ \ref{fig11}, Fig.\ \ref{fig12} shows the phase diagram associated with the singular behavior of the third-degree moments for the case $d=2$, $\sigma_1/\sigma_2=0.1$, and $m_1/m_2=4$. Here, as in Fig.\ \ref{fig11}, we have assumed that $\al_{22}=\al_{12}\equiv \al$. The curve $\al_c(a^*)$ splits the parameter space in two regions: the region above the curve corresponds to states $(\al, a^*)$ with finite (zero) values of these moments (i.e., $\ell_\text{min}>0$); the region below the curve provides states where those moments diverge in time. Thus, at a given value of $a^*$, there exists a critical value $\al_c(a^*)$ such that the moments are convergent for $\al> \al_c$. In particular,
we observe that $\al_c^*\to 0$ (and so, the moments become convergent) for sufficiently large values of the (reduced) shear rate $a^*$.

\section{Discussion}
\label{sec8}

It is well known that for molecular gases (i.e., particles colliding elastically), the model of Maxwell molecules
(namely, when the collision rate of two colliding particles is independent of their relative velocity) is a very useful starting point to obtain exactly transport properties in far from equilibrium states \cite{TM80,GS03}. On the other hand, when the collisions are inelastic and characterized by a constant coefficient of normal restitution $\al \leq 1$, one can also introduce the inelastic version of the Maxwell model (IMM). In this model, the form of the Boltzmann collision operator can be obtained from its corresponding form  for IHS by replacing the collision rate of hard spheres by an effective collision rate independent of the relative velocity. Thanks to this property, the collisional moments of the Boltzmann operator for IMM can be exactly written in terms of the velocity moments of the distributions $f_r$ and $f_s$ without knowing explicitly these distributions. This mathematical property of IMM opens up the possibility of obtaining exact results (the elastic limit $\al=1$ is a special limit) for granular flows, such as the Navier--Stokes transport coefficients \cite{S02,GA05} and/or the rheological properties of sheared granular gases \cite{G03,SG07,GT10}.

In the case of monocomponent granular gases, the choice of the Ikenberry polynomials $Y_{2p|\overline{q}}$  of degree $2p+q$ allows one to express the corresponding collisional moment $J_{2p|\overline{q}}$ as an eigenvalue $-\nu_{2p|q}$ times the velocity moment $M_{2p|\overline{q}}$ plus a bilinear combination of moments of degree less than $2p+q$. All the third and fourth degree collisional moments of IMM for monocomponent granular gases were evaluated in Ref.\  \cite{GS07}. We have extended in this paper the above results to the interesting case of binary granular mixtures. Due to the intricacy of the general problem, we have considered here situations where diffusion processes are absent. This means that the mean flow velocities $\mathbf{U}_r$ of each species are equal to the mean flow velocity $\mathbf{U}$ of the mixture ($\mathbf{U}_1=\mathbf{U}_2=\mathbf{U}$). Apart from this simplification, the results reported in this paper for the third and fourth degree collisional moments are exact for arbitrary values of the masses $m_r$, diameters $\sigma_r$, concentrations $x_r$, and coefficients of restitution $\al_{rs}$. In addition, all the derived expressions apply for any dimensionality $d$. Known results for three-dimensional molecular gases \cite{TM80,GS03} and for $d$-dimensional monocomponent granular gases \cite{GS07} are recovered. In the one-dimensional case ($d=1$) for binary granular mixtures, our results for the (isotropic) collisional moments $J_{2|0}^{(rs)}$ and $J_{4|0}^{(rs)}$ agree with the ones obtained by Marconi and Puglisi \cite{MP02a}. This shows the consistency of our general results with those previously reported in some particular limits.

As for monocomponent granular gases \cite{GS07}, we have observed that some of the eigenvalues $\nu_{2p|q}^{(rs)}$ exhibit a non-monotonic dependence on the coefficients of restitution $\al_{rs}$ at given values of the mass and diameter ratios and the concentration. We have also seen that the impact of the inelasticity in collisions on the eigenvalues is in general important, specially in the case of the eigenvalues associated with the self-collision terms. Although the above observations are restricted to the moments of degree $2p+q\leq 4$, we expect that they extend to moments of higher degree.

The knowledge of the second, third, and fourth degree collisional moments for inelastic Maxwell mixtures opens up the possibility of studying specific nonequilibrium situations. We have analyzed in this paper two different problems. First, we have studied the time evolution of the moments of degree equal to or less than 4 in the HCS. In this state, given that the granular temperature $T$ decreases in time, one has to scale the moments with the thermal speed $v_0(t)=\sqrt{2T(t)(m_1+m_2)/m_1 m_2}$ to reach steady values in the long time limit. Our analysis shows that while all the second degree moments tend towards finite values for long times, the third degree moments $M_{2|i}^{*(r)}$ (which are related to the heat flux) can diverge in a region of the parameter space of the mixture. This sort of divergence also appears in all (isotropic and anisotropic) fourth degree moments. The above conclusions contrast with the ones achieved for monocomponent granular gases \cite{GS07} where all the moments of degree $2p+q\leq 4$ are convergent for $d\geq2$. The singular behavior of the third degree moments is consistent with an algebraic high velocity tail of the form $f_r(V)\sim V^{-(d+s)}$, where $s\leq 3$ when the moments $M_{2|i}^{*(r)}$ are divergent. We plan to explore this possibility in a forthcoming work.

As a second application, we have analyzed the time evolution of the second and third degree moments of a \emph{sheared} granular binary mixture where one of the species is present in tracer concentration. In this situation, given that the dynamic properties of the excess species coincide with those previously obtained for simple granular gases \cite{SG07}, the study is focused on the tracer species. In particular, in contrast to the findings of monocomponent granular gases of IMM \cite{SG07}, our results show that the (scaled) third-degree moments of the tracer species can diverge in time for given values of the parameters of the mixture. This is the expected result according to the analysis made in the HCS. However, it is quite apparent that in general those moments become convergent for sufficiently large values of the (reduced) shear rate. Thus, one can conclude that the main effect of the shear rate on the third-degree moments of tracer species is to increase the size of the region where those moments are convergent.

One of the limitations of the results derived in this paper is its restriction to non-equilibrium situations where the flow velocities of both species are equal ($\mathbf{U}_1=\mathbf{U}_2$). This yields a vanishing mass flux ($\mathbf{j}_r=\mathbf{0}$). The extension to situations where $\mathbf{U}_1\neq \mathbf{U}_2$ is possible but the determination of these new terms (coupling $\mathbf{j}_r$ with other moments) in the corresponding collisional moments involves a quite long and tedious calculation. A previous work \cite{GA05} on IMM has accounted for these new contributions for the collisional moments $J_{0|i}^{(rs)}$, $J_{2|0}^{(rs)}$, $J_{0|ij}^{(rs)}$, and $J_{2|i}^{(rs)}$. We plan to extend the present expressions for the collisional moments $J_{0|ijk}^{(rs)}$, $J_{4|0}^{(rs)}$, $J_{2|ij}^{(rs)}$, and $J_{0|ijk\ell}^{(rs)}$ for non-vanishing mass fluxes in the near future. This will allow us to obtain the collisional moments of second, third and fourth degree in a granular binary mixture of IMM without any kind of restriction.

The fact that the third and fourth degree moments in the HCS may be divergent have important physical consequences on the transport coefficients given that the HCS plays the role of the reference state in the Chapman--Enskog perturbative solution \cite{CC70} to the Boltzmann equation. In particular, as Brey \emph{et al}. \cite{BGM10} pointed out in the monodisperse case, the transport coefficients associated with the heat flux can be divergent for values of $\al<\al_c$ ($\al_c=\frac{1}{3}$ at $d=2$ and $\al_c=\frac{1}{9}$ at $d=3$). These authors \cite{BGM10} found that below the critical value $\al_c$, one of the kinetic modes (the one associated with the heat flux) decays more slowly than the hydrodynamic mode associated with the granular temperature. They concluded that a hydrodynamic description is not possible for values of $\al<\al_c$. A similar behavior is expected for granular mixtures, although the values of $\al_{rs,c}$ will have a complex dependence on the concentration and the mass and diameter ratios. Regarding the above point, it is interesting to remark that a slightly different view to the one offered in Ref.\ \cite{BGM10} on the singular behavior of the heat flux transport coefficients has been provided in Ref.\ \cite{GS11}. According to this work, the origin of the above divergence could be also associated with the possible high-velocity tail of the first-order distributions $f_r^{(1)}$ of the Chapman--Enskog solution. Thus, although $f_r^{(1)}$ could be well defined for any value of the coefficients of restitution, its third-order velocity moments (such as the heat flux) might diverge due to the high-velocity tail of this distribution. In any case and according to the results reported in the present paper for the velocity moments in the HCS for granular mixtures, given that the critical values $\al_{rs,c}$ are in general small, the possible breakdown of granular hydrodynamics has no important consequences for practical purposes.

The explicit results provided in this paper can be employed to analyze different nonequilibrium problems. As mentioned before, one of them is to extend our analysis to binary mixtures with arbitrary values of the concentration. In the USF problem, apart from the rheological properties \cite{G03,GT10}, it would be interesting to study the time evolution of the fourth degree velocity moments towards their steady values and investigate whether these moments can be divergent as occurs for elastic collisions \cite{SG95}. Another interesting application of the present results is to determine some of the generalized transport coefficients characterizing small perturbations around the simple shear flow problem \cite{GT15,G07}. Work along these lines will be carried out in the near future.




\section*{Acknowledgments}

The authors  acknowledge financial support from Grant PID2020-112936GB-I00 funded by MCIN/AEI/ 10.13039/501100011033 (V. G.), and from Grants IB20079 (V. G.) and GR21014 (C. S. R. and V. G.) funded by Junta de Extremadura (Spain) and by ERDF ``A way of making Europe.''

\appendix
\section{Some technical details in the evaluation of the collisional moments}
\label{appA}

In this Appendix we give some technical details on the derivation of the collisional moments $J_{2p|i_1i_2\ldots i_q}^{(rs)}$ associated with the Ikenberry polynomials of third and fourth degree when the mass flux of each species vanishes.

We consider for the sake of concreteness the (anisotropic) third degree collisional moment
\begin{eqnarray}
\label{a1}
J_{0|ijk}^{(rs)}&=&\int d{\bf V}\left[V_i V_j V_k-
\frac{1}{d+2}V^2\left(V_i\delta_{jk}+V_j\delta_{ik}+V_k
\delta_{ij}\right)\right]J_{rs}[f_r,f_s]\nonumber\\
&=&J_{ijk}^{(rs)}-\frac{1}{d+2}\left(J_{2|i}^{(rs)}\delta_{jk}+J_{2|j}^{(rs)}\delta_{ik}+J_{2|k}^{(rs)}\delta_{ij}\right),
\end{eqnarray}
where
\begin{equation}
\label{a2}
J_{ijk}^{(rs)}=\int d{\bf V} V_iV_jV_k J_{rs}[f_r,f_s],
\end{equation}
\begin{equation}
\label{a3}
J_{2|i}^{(rs)}=\int d{\bf V} V^2V_i J_{rs}[f_r,f_s].
\end{equation}
Let us evaluate first the third degree (canonical) collisional moment $J_{ijk}^{(rs)}$. Taking into account the property (\ref{1.3.1}), Eq.\ \eqref{a2} can be written as
\begin{equation}
\label{a6}
J_{ijk}^{(rs)}=
\frac{\omega_{rs}}{n_s\Omega_d}
\int \,d{\bf V}_{1}\,\int \,d{\bf V}_{2}f_r({\bf V}_{1})f_s({\bf V}_{2})
\int d\widehat{\boldsymbol{\sigma}}\,\left(V_{1i}' V_{1j}' V_{1k}'-
V_{1i} V_{1j} V_{1k}\right),
\end{equation}
where Eq.\ \eqref{1.3.2} gives the relationship between the post-collisional velocity $\mathbf{V}_1'$ and the pre-collisional velocity $\mathbf{V}_1$. The fact that ${\bf U}_1={\bf U}_2={\bf U}$ allows one to map some of the results obtained for monodisperse gases \cite{GS07} by making the change
\begin{equation}
\label{a5.1}
\alpha \to \beta_{rs}\equiv 2\mu_{sr}(1+\alpha_{rs})-1.
\end{equation}
In particular, the scattering rule (\ref{1.3.1}) implies that
\begin{eqnarray}
\label{a7}
V_{1i}'V_{1j}'V_{1k}'- V_{1i} V_{1j} V_{1k}&=&
-\frac{1+\beta_{rs}}{2}(\widehat{\boldsymbol{\sigma }}\cdot {\bf
g})\Big\{V_{1i}V_{1j}\widehat{\sigma}_k+V_{1i}V_{1k}\widehat{\sigma}_j+V_{1j}V_{1k}\widehat{\sigma}_i
\nonumber\\
& &
-\frac{1+\beta_{rs}}{2}(\widehat{\boldsymbol{\sigma}}\cdot {\bf
g})\Big[V_{1i}\widehat{\sigma}_j\widehat{\sigma}_k+
V_{1j}\widehat{\sigma}_i\widehat{\sigma}_k+V_{1k}\widehat{\sigma}_i\widehat{\sigma}_j\nonumber\\
& &
-\frac{1+\beta_{rs}}{2}(\widehat{\boldsymbol{\sigma }}\cdot {\bf g})
\widehat{\sigma}_i\widehat{\sigma}_j\widehat{\sigma}_k\Big]\Big\}.\end{eqnarray}
To carry out the angular integrations in (\ref{a6}) one needs the results
\begin{equation}
\label{a8} \int d\widehat{\boldsymbol{\sigma}}\,
(\widehat{\boldsymbol{\sigma}}\cdot
{\bf g})^{2k+1} \widehat{{\sigma}}_i=B_{k+1} g^{2k} g_{i},
\end{equation}
\begin{equation}
\label{a9} \int d\widehat{\boldsymbol{\sigma}}\,(\widehat{\boldsymbol{\sigma}}\cdot
{\bf g})^{2k} \widehat{{\sigma}}_i\widehat{{\sigma}}_j =\frac{B_{k}}{2k+d}g^{2(k-1)}
\left(2k g_{i}g_{j}+g^2 \delta_{ij}\right),
\end{equation}
\beqa
\label{a10} \int d\widehat{\boldsymbol{\sigma}}\,
(\widehat{\boldsymbol{\sigma}}\cdot
{\bf g})^{2k+1}
\widehat{\sigma}_i\widehat{\sigma}_j\widehat{\sigma}_\ell&=&
\frac{B_{k+1}}{2(k+1)+d
}g^{2(k-1)}
\big[ 2k g_{i}g_{j}g_{\ell}+g^2
\big(\delta_{ij}g_{\ell}+\delta_{i\ell}g_{j} \nonumber\\
& &
+\delta_{j\ell}g_{i}\big)\big].
\eeqa
Here, the coefficients $B_k$ are \cite{NE98}
\begin{equation}
\label{a11} B_{k}=\int
d\widehat{\boldsymbol{\sigma}}\,(\widehat{\boldsymbol{\sigma}}\cdot {\widehat{\bf
g}})^{2k}=\Omega_d\pi^{-1/2}\frac{\Gamma\left(\frac{d}{2}\right)
\Gamma\left(k+\frac{1}{2}\right)}{\Gamma\left(k+\frac{d}{2}\right)}.
\end{equation}
Making use of Eqs.\ (\ref{a7})--(\ref{a10}), one gets
\begin{eqnarray}
\label{a12}
J_{ijk}^{(rs)}&=&-\frac{1}{2d(d+2)}\frac{\omega_{rs}}{n_s}(1+\beta_{rs})
\Big\{(d+2)
\langle V_{1i} V_{1j} g_{k}+V_{1i} V_{1k} g_j+
V_{1j} V_{1k} g_i \rangle\nonumber\\
& & -\frac{(1+\beta_{rs})}{2}\big[2\langle g_i g_j V_{1k}+g_i g_k V_{1j}+
g_j g_k V_{1i} \rangle \nonumber\\
& & + \langle g^2(V_{1i}\delta_{jk}
+V_{1j}\delta_{ik}+V_{1k}\delta_{ij})\rangle\big]+\frac{3}{4(d+4)}
(1+\beta_{rs})^2\nonumber\\
& & \times
\big[2\langle g_{i} g_{j} g_{k}+g^2(g_{i}\delta_{jk}
+g_{j}\delta_{ik}+g_{k}\delta_{ij})\rangle\big]
\Big\},
\end{eqnarray}
where the brackets are defined as
\begin{equation}
\label{a13}
\langle h({\bf V}_1,{\bf V}_2)\rangle \equiv \int
d{\bf V}_1\int d{\bf V}_2 h({\bf V}_1,{\bf V}_2) f_r({\bf V}_1)f_s({\bf V}_2).
\end{equation}
The integrations over velocities give the relations
\begin{equation}
\label{a14}
\langle V_{1i} V_{1j} V_{1k}\rangle=\langle g_{i} g_{j} V_{1k}\rangle
=n_s M_{ijk}^{(r)},
\quad
\langle g^2V_{1i}\rangle=n_s M_{2|i}^{(r)},
\end{equation}
\begin{equation}
\label{a15}
\langle g_{i} g_{j} g_{k}\rangle=n_s M_{ijk}^{(r)}-n_r M_{ijk}^{(rs)},
\end{equation}
where
\begin{equation}
\label{a16}
M_{ijk}^{(r)}=\int\; d{\bf V} V_{i} V_{j} V_{k}  f_r({\bf V}).
\end{equation}
Therefore, from Eqs. \ (\ref{a12}) and (\ref{a14})--(\ref{a16}) one finally obtains
\begin{eqnarray}
\label{a17}
J_{ijk}^{(rs)}&=&-\frac{3}{4d(d+2)(d+4)}\frac{\omega_{rs}}{n_s}
(1+\beta_{rs})
\Big\{ [\beta_{rs}^2-2(d+3)\beta_{rs}+2d^2+10d+9)]\nonumber\\
& & \times
n_sM_{ijk}^{(r)}
-(1+\beta_{rs})^2n_rM_{ijk}^{(s)}
\nonumber\\
& & +\frac{(1+\beta_{rs})}{6}(3\beta_{rs}-2d-5)n_s\left(M_{2|i}^{(r)}\delta_{kj}+
M_{2|j}^{(r)}\delta_{ik}+M_{2|k}^{(r)}\delta_{ij}\right)\nonumber\\
& & -\frac{(1+\beta_{rs})^2}{2}n_r\left(M_{2|i}^{(s)}\delta_{jk}+
M_{2|j}^{(s)}\delta_{ik}+M_{2|k}^{(s)}\delta_{ij}\right)
\Big\}.
\end{eqnarray}
If one makes $j=k$ and sum over $j$ one obtains the expression (\ref{1.13}) for $J_{2|i}^{(rs)}$. Also, by subtracting $\left(J_{2|i}^{(rs)}\delta_{jk}+J_{2|i}^{(rs)}\delta_{jk}+J_{2|i}^{(rs)}\delta_{jk}\right)/(d+2)$ from both sides of Eq.\ (\ref{a17}) one gets Eq.\ (\ref{1.14}) for $J_{0|ijk}^{(rs)}$.

The calculations for the fourth degree collisional moments are similar to those carried out for the third degree moments. In addition, most of the mathematical steps followed to get them can be easily mapped from those made for monodisperse gases in Ref.\ \cite{GS07} by replacing $\alpha\to \beta_{rs}$. After a long and tedious algebra one obtains the results displayed along the Section \ref{sec2}.

\section{Explicit forms of the eigenvalues}
\label{appB}

In this Appendix we give the explicit expressions of the eigenvalues $\nu_{2p|q}^{(11)}$
and $\nu_{2p|q}^{(12)}$ with $2p+q \leq 4$ of the collisional moments $J_{2p|\bar{q}}^{*(rs)}$. In the case of the second degree collisional moments, the eigenvalues are
\begin{equation}
\label{b1}
\nu_{2|0}^{(11)}=\omega_{11}^* \frac{1-\alpha_{11}^2}{2d}+
\frac{\omega_{12}^*}{4d}(1+\beta_{12})(3-\beta_{12}), \quad \nu_{2|0}^{(12)}=-\frac{\omega_{12}^*}{4d}(1+\beta_{12})^2,
\end{equation}
\begin{equation}
\label{b2}
\nu_{0|ij}^{(11)}=\frac{\omega_{11}^*}{d(d+2)} (1+\alpha_{11})(d+1-\alpha_{11})+
\frac{\omega_{12}^*}{2d(d+2)}(1+\beta_{12})(2d+3-\beta_{12}),
\end{equation}
\beq
\label{b2.0}
\nu_{0|ij}^{(12)}=-\frac{\omega_{12}^*}{2d(d+2)}(1+\beta_{12})^2,
\eeq
where $\omega_{rs}^*=\omega_{rs}/\nu_0$. The corresponding eigenvalues in the case of the third degree collisional moments are given by
\beqa
\label{b3}
\nu_{2|i}^{(11)}&=&\frac{\omega_{11}^*}{4d(d+2)} (1+\alpha_{11})[4+5d-(d+8)\alpha_{11}]+
\frac{\omega_{12}^*}{8d(d+2)}(1+\beta_{12})\nonumber\\
& & \times
[3\beta_{12}^2-2(d+5)\beta_{12}+10d+11],
\eeqa
\begin{equation}
\label{b4}
\nu_{2|i}^{(12)}=-3\frac{\omega_{12}^*}{8d(d+2)}(1+\beta_{12})^3,
\end{equation}
\beqa
\label{b4.0}
\nu_{0|ijk}^{(11)}&=&\frac{3}{2}\frac{\omega_{11}^*}{d(d+2)}(1+\alpha_{11})
[d+1-\alpha_{11}]+
3\frac{\omega_{12}^*}{4d(d+2)(d+4)}(1+\beta_{12})\nonumber\\
& & \times [\beta_{12}^2-2(d+3)
\beta_{12}+2d^2+10d+9],
\eeqa
\beq
\label{b4.1}
\nu_{0|ijk}^{(12)}=-3\frac{\omega_{12}^*}{4d(d+2)(d+4)}(1+\beta_{12})^3.
\eeq
Finally, the eigenvalues of the fourth degree collisional moments are
\beqa
\label{b5}
\nu_{4|0}^{(11)}&=&\frac{\omega_{11}^*}{8d(d+2)} (1+\alpha_{11})[9+12d-(4d+17)\alpha_{11}+3\alpha_{11}^2-
3\alpha_{11}^3]\nonumber\\
& & +
\frac{\omega_{12}^*}{16d(d+2)}(1+\beta_{12})
(3-\beta_{12})[3\beta_{12}^2-6\beta_{12}+8d+7],
\eeqa
\begin{equation}
\label{b6}
\nu_{4|0}^{(12)}=-3\frac{\omega_{12}^*}{16d(d+2)}(1+\beta_{12})^4,
\end{equation}
\begin{eqnarray}
\label{b7}
\nu_{2|ij}^{(11)}&=&\frac{\omega_{11}^*}{4d(d+2)(d+4)} (1+\alpha_{11})\big[7d^2+31d+18-(d^2+14d+34)\alpha_{11}\nonumber\\
& & +3(d+2)\alpha_{11}^2
-6\alpha_{11}^3\big]+
\frac{\omega_{12}^*}{4d(d+2)(d+4)}(1+\beta_{12})\nonumber\\
& & \times
\big[7d^2+31d+21-(d^2+14d+25)\beta_{12}+3(d+5)\beta_{12}^2
-3\beta_{12}^3\big],
\end{eqnarray}
\begin{equation}
\label{b8}
\nu_{2|ij}^{(12)}=-3\frac{\omega_{12}^*}{4d(d+2)(d+4)}(1+\beta_{12})^4,
\end{equation}
\begin{eqnarray}
\label{b9}
\nu_{0|ijk\ell}^{(11)}&=&\frac{\omega_{11}^*}{d(d+2)(d+4)(d+6)} (1+\alpha_{11})\big[2d^3+21d^2+61d+39 \nonumber\\
& &
-3(d+3)(d+5)\alpha_{11}+3(d+3)\alpha_{11}^2-3\alpha_{11}^3\big]
\nonumber\\
& &+
\frac{\omega_{12}^*}{2d(d+2)(d+4)(d+6)}(1+\beta_{12})
\big[4d^3+42d^2+122d+81 \nonumber\\
& & -3(2d^2+16d+27)\beta_{12}+(6d+27)\beta_{12}^2
-3\beta_{12}^3\big],\nonumber\\
\end{eqnarray}
\begin{equation}
\label{b10}
\nu_{0|ijk\ell}^{(12)}=-3\frac{\omega_{12}^*}{2d(d+2)(d+4)(d+6)}(1+\beta_{12})^4.
\end{equation}

\end{document}